\documentclass[twoside,twocolumn,9pt]{article}
\usepackage{amsmath}
\usepackage{amssymb}
\usepackage{amsfonts}
\usepackage{extsizes}
\usepackage[super,sort&compress,comma]{natbib} 
\usepackage[version=3]{mhchem}
\usepackage[left=1.5cm, right=1.5cm, top=1.785cm, bottom=2.0cm]{geometry}
\usepackage{balance}
\usepackage{times,mathptmx}
\usepackage{sectsty}
\usepackage{graphicx} 
\usepackage{lastpage}
\usepackage[format=plain,justification=justified,singlelinecheck=false,font={stretch=1.125,small,sf},labelfont=bf,labelsep=space]{caption}
\usepackage{float}
\usepackage{fancyhdr}
\usepackage{fnpos}
\usepackage[english]{babel}
\addto{\captionsenglish}{%
  
}
\usepackage{array}
\usepackage{droidsans}
\usepackage{charter}
\usepackage[T1]{fontenc}
\usepackage[usenames,dvipsnames]{xcolor}
\usepackage{setspace}
\usepackage[compact]{titlesec}
\usepackage{hyperref}
\usepackage{epstopdf}%This line makes .eps figures into .pdf - please comment out if not required.

\definecolor{cream}{RGB}{222,217,201}
\usepackage{xcolor}
\usepackage{tikz,lipsum,lmodern}
\usepackage[most]{tcolorbox}

\begin{document}

\pagestyle{fancy}
\thispagestyle{plain}
\fancypagestyle{plain}{

%%%HEADER%%%
\fancyhead[C]{\includegraphics[width=18.5cm]{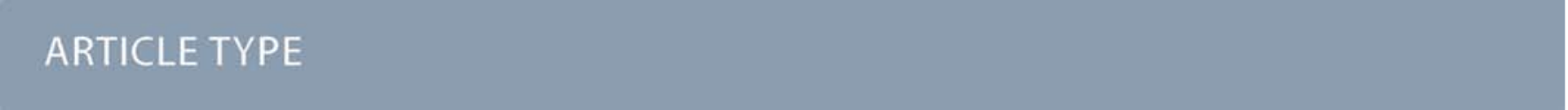}}
\fancyhead[L]{\hspace{0cm}\vspace{1.5cm}\includegraphics[height=30pt]{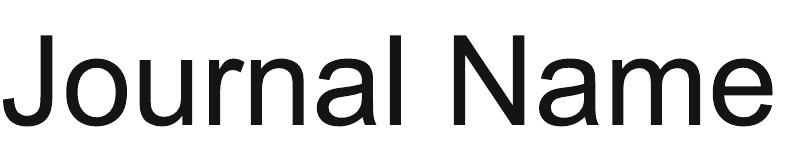}}
\fancyhead[R]{\hspace{0cm}\vspace{1.7cm}\includegraphics[height=55pt]{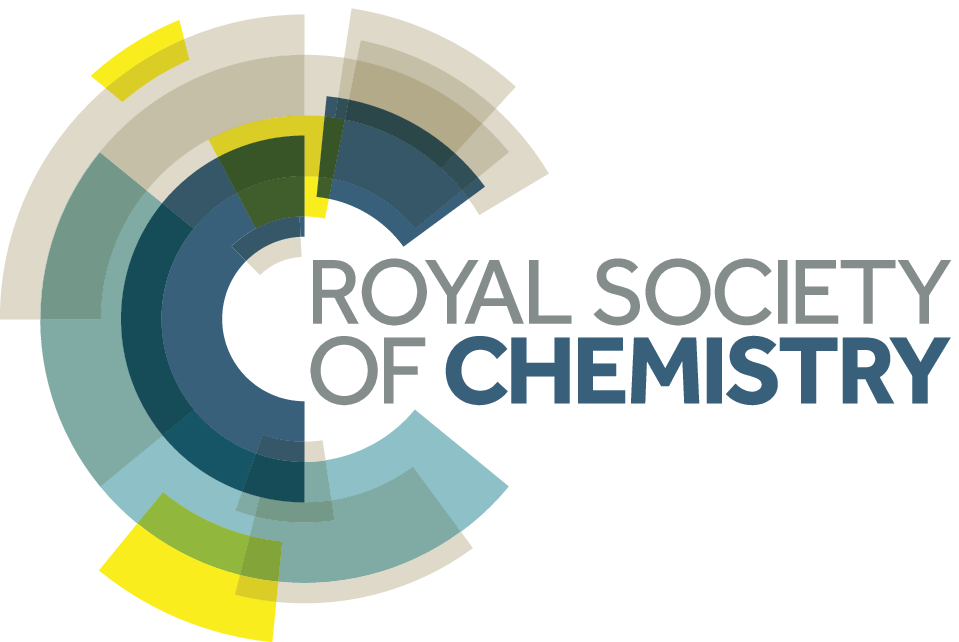}}
\renewcommand{\headrulewidth}{0pt}
}
%%%END OF HEADER%%%

%%%PAGE SETUP - Please do not change any commands within this section%%%
\makeFNbottom
\makeatletter
\renewcommand\LARGE{\@setfontsize\LARGE{15pt}{17}}
\renewcommand\Large{\@setfontsize\Large{12pt}{14}}
\renewcommand\large{\@setfontsize\large{10pt}{12}}
\renewcommand\footnotesize{\@setfontsize\footnotesize{7pt}{10}}
\makeatother

\renewcommand{\thefootnote}{\fnsymbol{footnote}}
\renewcommand\footnoterule{\vspace*{1pt}% 
\color{cream}\hrule width 3.5in height 0.4pt \color{black}\vspace*{5pt}} 
\setcounter{secnumdepth}{5}

\makeatletter 
\renewcommand\@biblabel[1]{#1}            
\renewcommand\@makefntext[1]% 
{\noindent\makebox[0pt][r]{\@thefnmark\,}#1}
\makeatother 
\renewcommand{\figurename}{\small{Fig.}~}
\sectionfont{\sffamily\Large}
\subsectionfont{\normalsize}
\subsubsectionfont{\bf}
\setstretch{1.125} %In particular, please do not alter this line.
\setlength{\skip\footins}{0.8cm}
\setlength{\footnotesep}{0.25cm}
\setlength{\jot}{10pt}
\titlespacing*{\section}{0pt}{4pt}{4pt}
\titlespacing*{\subsection}{0pt}{15pt}{1pt}
%%%END OF PAGE SETUP%%%

%%%FOOTER%%%
\fancyfoot{}
\fancyfoot[LO,RE]{\vspace{-7.1pt}\includegraphics[height=9pt]{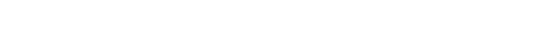}}
\fancyfoot[CO]{\vspace{-7.1pt}\hspace{13.2cm}\includegraphics{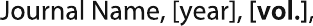}}
\fancyfoot[CE]{\vspace{-7.2pt}\hspace{-14.2cm}\includegraphics{head_foot/RF}}
\fancyfoot[RO]{\footnotesize{\sffamily{1--\pageref{LastPage} ~\textbar  \hspace{2pt}\thepage}}}
\fancyfoot[LE]{\footnotesize{\sffamily{\thepage~\textbar\hspace{3.45cm} 1--\pageref{LastPage}}}}
\fancyhead{}
\renewcommand{\headrulewidth}{0pt} 
\renewcommand{\footrulewidth}{0pt}
\setlength{\arrayrulewidth}{1pt}
\setlength{\columnsep}{6.5mm}
\setlength\bibsep{1pt}
%%%END OF FOOTER%%%

%%%FIGURE SETUP - please do not change any commands within this section%%%
\makeatletter 
\newlength{\figrulesep} 
\setlength{\figrulesep}{0.5\textfloatsep} 

\newcommand{\topfigrule}{\vspace*{-1pt}% 
\noindent{\color{cream}\rule[-\figrulesep]{\columnwidth}{1.5pt}} }

\newcommand{\botfigrule}{\vspace*{-2pt}% 
\noindent{\color{cream}\rule[\figrulesep]{\columnwidth}{1.5pt}} }

\newcommand{\dblfigrule}{\vspace*{-1pt}% 
\noindent{\color{cream}\rule[-\figrulesep]{\textwidth}{1.5pt}} }

\makeatother
%%%END OF FIGURE SETUP%%%

%%%TITLE, AUTHORS AND ABSTRACT%%%
\twocolumn[
  \begin{@twocolumnfalse}
\vspace{3cm}
\sffamily
\begin{tabular}{m{4.5cm} p{13.5cm}}

\includegraphics{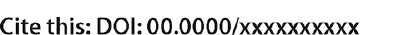} & \noindent\LARGE{\textbf{Multidimensional coherent  spectroscopy of excitons in $\pi$-conjugated polymer}} \\%Article title goes here instead of the text "This is the title"
% & \noindent\LARGE{\textbf{in $\pi$-conjugated polymers}} \\%Article title goes here instead of the text "This is the title"
\vspace{0.3cm} & \vspace{0.3cm} \\

& \noindent\large{Elizabeth~Guti\'errez-Meza,\textit{$^{a}$$^{\dag}$} Alejandro~Vega-Flick,\textit{$^{a}$$^{\dag}$}  
Eric~R.~Bittner,\textit{$^{b,c}$}} and \\
& \noindent\large{Carlos~Silva-Acu\~na\textit{$^{a,d,e}$$^{\ast}$}} \\%Author names go here instead of "Full name", etc.
 %& \noindent\large{Full Name,$^{\ast}$\textit{$^{a}$} Full Name,\textit{$^{b\ddag}$} and Full Name\textit{$^{a}$}} \\%Author names go here instead of "Full name", etc.

\includegraphics{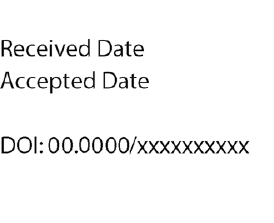} & \noindent\normalsize{
%The photophysics of excitons in $\pi$-conjugated polymers has been of considerable interest to the scientific community because of the fundamental role they play in charge and energy transfer processes. These excitons can be optically probed via spectroscopy methods, providing a wealth of information regarding polymer molecular structure and its energy landscape. Nonetheless, there are intrinsic limitations in the information that can be obtained with linear spectroscopy compared to non-linear coherent techniques. Examples of this are separation of homogeneous and inhomogeneous broadening mechanisms, population dynamics, exciton-vibrational coupling, etc. In the present work we discuss recent studies of non-linear two-dimensional spectroscopy techniques on $\pi$-conjugated polymer:fullerene blends composed of P3HT:PCBM. We explore ... 

The photophysics of $\pi$-conjugated polymers has been of considerable interest over the last three decades because of their organic semiconductor properties. Primary photoexcitations, Frenkel excitons, can be probed optically by means of numerous linear spectroscopies, providing a wealth of information on the the strength of excitonic coupling,  the exciton coherence length, and on the nature of the disordered energy landscape. Nonetheless, there are intrinsic limitations in the information that can be obtained with linear spectroscopy compared to non-linear coherent techniques. Examples of this are the separation of homogeneous and inhomogeneous broadening contributions to the total exciton spectral lineshape, the detailed spectral structure of exciton-vibrational coupling, and correlations between optical excitations, including exciton-charge and exciton-exciton correlations. In the present work, we discuss the role of two-dimensional coherent excitation spectroscopy and review its applications towards unravelling the basic photophysics of $\pi$-conjugated polymers as well as polymer:fullerene blends, and we argue that these techniques are now valuable mainstream materials optical probes. 

%to establish the dominant character of the disordered energy landscape either if it is static or dynamic

} \\

\end{tabular}

 \end{@twocolumnfalse} \vspace{0.6cm}
  ]
%%%END OF TITLE, AUTHORS AND ABSTRACT%%%

%%%FONT SETUP - please do not change any commands within this section
\renewcommand*\rmdefault{bch}\normalfont\upshape
\rmfamily
\section*{}
\vspace{-1cm}

%%%FOOTNOTES%%%
%\footnotetext{\textit{$^{a}$~Address, Address, Town, Country. Fax: XX XXXX XXXX; Tel: XX XXXX XXXX; E-mail: xxxx@aaa.bbb.ccc}}
\footnotetext{\textit{$^{a}$~School of Chemistry and Biochemistry, Georgia Institute of Technology, 901 Atlantic Drive NW, Atlanta, Georgia 30332, United~States. E-mail: carlos.silva@gatech.edu}}
\footnotetext{\textit{$^{b}$~Department of Chemistry, University of Houston, Houston, Texas 77204, United~States}}
\footnotetext{\textit{$^{c}$~Center for Non-Linear Studies, Los Alamos National Lab, Los Alamos, New Mexico 87545, United States}}
\footnotetext{\textit{$^{d}$~School of Physics, Georgia Institute of Technology, 837 State Street NW, Atlanta, Georgia 30332, United~States }}
\footnotetext{\textit{$^{e}$~School of Materials Science and Engineering, Georgia Institute of Technology, North Avenue, Atlanta, GA 30332, United~States }}
\footnotetext{\dag~A.V.F. and E.G.M.\ are first co-authors of this manuscript.}
%
%Please use \dag to cite the ESI in the main text of the article.
%If you article does not have ESI please remove the the \dag symbol from the title and the footnotetext below.
%\footnotetext{\dag~Electronic Supplementary Information (ESI) available: [details of any supplementary information available should be included here]. See DOI: 00.0000/00000000.}
%additional addresses can be cited as above using the lower-case letters, c, d, e... If all authors are from the same address, no letter is required

%%%END OF FOOTNOTES%%%

\begin{tcolorbox}[colback=gray!20!white,colframe=black!75!black,arc=0mm,boxrule=0.1mm]
\subsection*{10$^{\mathrm{th}}$ Anniversary Statement}
%The Journal of Materials Chemistry was split into three independent journals in 2013, meaning that in 2023, Journal of Materials Chemistry A, B and C will be celebrating ten years of publication. Please include an anniversary statement here, of no more than 200 words, to highlight your relationship to the journal. For example, this could include your thoughts on the journal and its community, reflections on your previous work published in the journal or future directions in the field.
The influence of the \textit{Journal of Materials Chemistry C} in the development of the basic science of conjugated polymers is substantial, and C.S.A.\ has published some of his key contributions to the photophysics of these materials in the journal. It is his firm conviction that advanced optical probes are now at the point of maturity that they are to be considered as valuable techniques within the materials characterization toolbox, much like ``standard'' techniques such as linear absorption and emission techniques. The readership of the journal is broad, spanning all aspects of research on conjugated polymers, ranging from synthesis, processing, physical characterization, and engineering of their technologies, and this contribution to the 10$^{\mathrm{th}}$ Anniversary Issue is aimed at the entire community. The authors therefore hope to stimulate further use of nonlinear coherent spectroscopies in the research of excitonic properties of conjugated polymer materials and their devices by connecting the ultrafast spectroscopy and materials science communities. 
\end{tcolorbox}

%%%MAIN TEXT%%%%

\section{Introduction}

\begin{figure*}[ht]
    \centering
    \includegraphics[width=\textwidth]{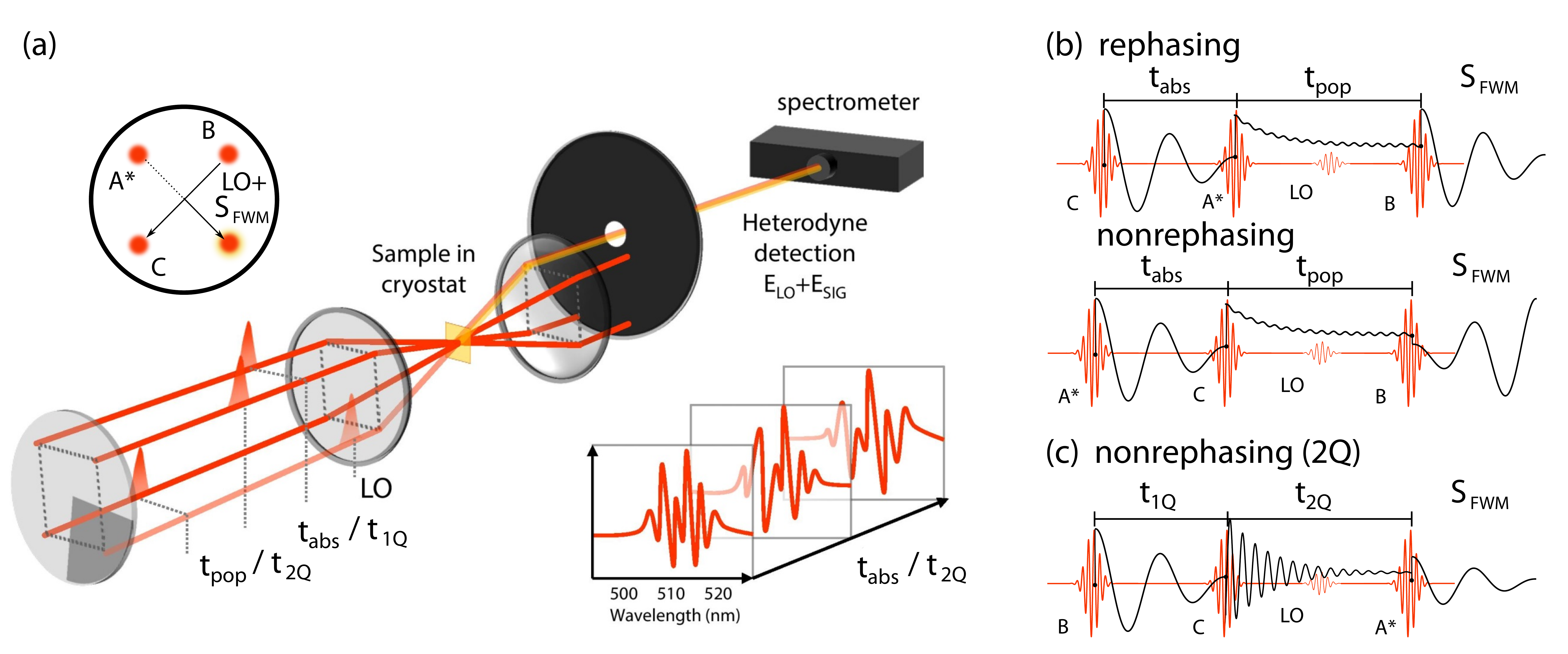}
    \caption{Schematic representation of a two-dimensional coherent spectroscopy experimental arrangement. The geometry of the excitation pulse-train beam pattern (red) and the resonant four-wave mixing signal ($S_{\mathrm{FWM}}$, yellow-orange), detected by interference with a local oscillator (LO), is shown in (a). This scheme uses a BOXCARS beam geometry, in which three pulse trains (A, B, C) propagating along the corners of a square are focused onto the sample with a common lens, defining incident wavevector $\Vec{k}_A$, $\Vec{k}_B$, and $\Vec{k}_C$. The LO beam, on the fourth apex of the incident beam geometry, co-propagates with $S_{\mathrm{FWM}}$ with wavevector imposed by the chosen phase matching conditions. The spectral interferogram of $S_{\mathrm{FWM}}$ and the LO beam is recorded at every time step. (b) By controlling the order of the pulse sequence with this beam geometry, we measure two distinct $S_{\mathrm{FWM}}$ responses: the \emph{rephasing} signal with wavevector $\Vec{k}_{S_{\mathrm{FWM}}} = \Vec{k}_B - \Vec{k}_A + \Vec{k}_C$ and \emph{nonrephasing} signal with wavevector $\Vec{k}_{S_{\mathrm{FWM}}} = \Vec{k}_A - \Vec{k}_B + \Vec{k}_C$. The total nonlinear response is then given by the sum of the rephasing and non-rephasing spectra. (c) Pulse sequence for two-quantum (2Q) coherence measurement.  Figure taken with permission from Ref.~54.}
    \label{fig:colbert}
\end{figure*}
\begin{figure*}[ht]
    \centering
    \includegraphics[width=0.8\textwidth]{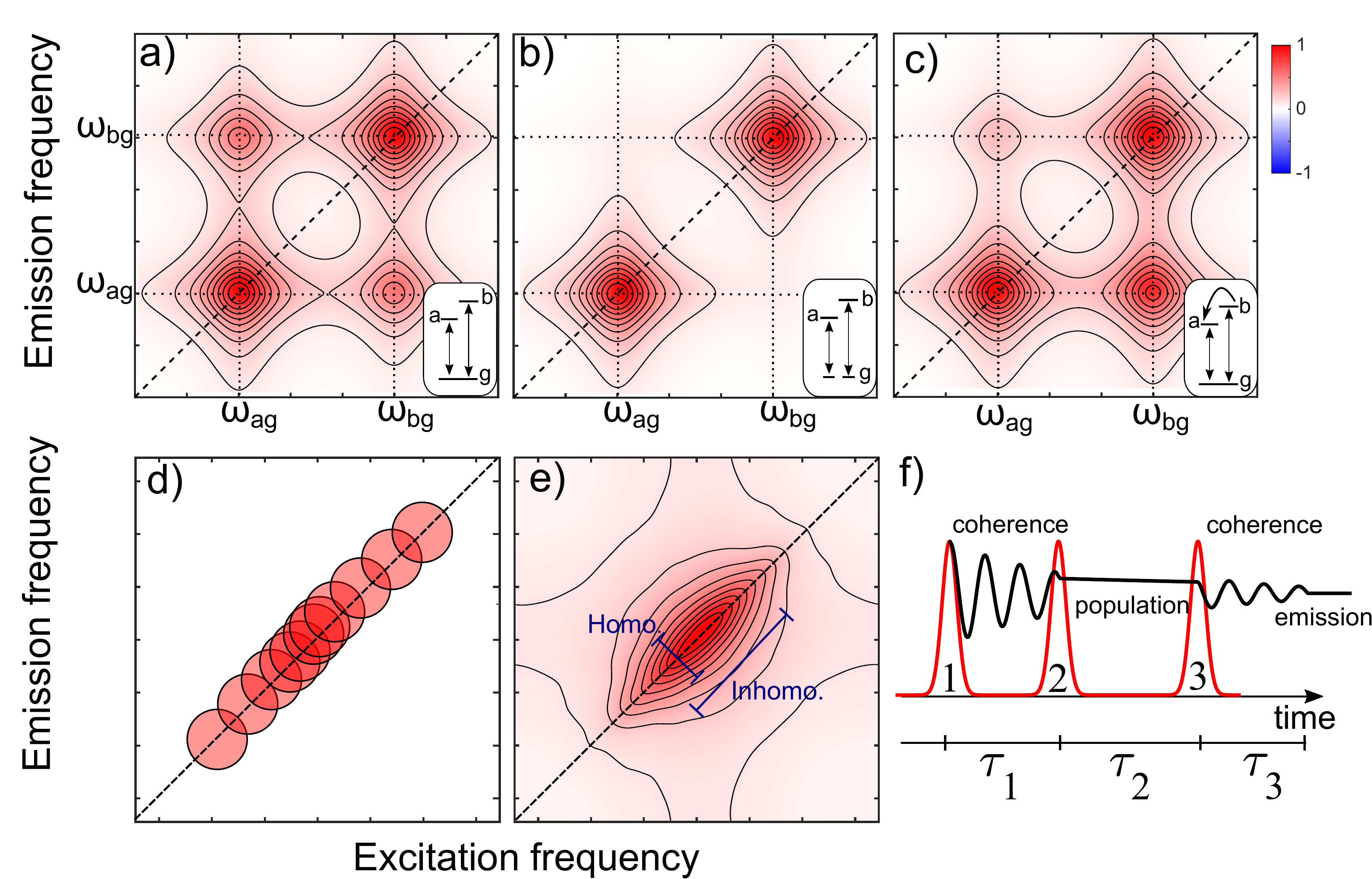}
    \caption{Example of a 2D spectra of a system with two excited levels $|a \rangle$, $|b \rangle$ and a ground state $|g \rangle$. (a) The spectrum is symmetric along the diagonal since the states share a common ground, (b) the states are decoupled, (c) coupling of the b and a states leads to relaxation of b towards a, which translates to an asymmetric amplitude of the off-diagonal peaks. Insets show the energy diagrams of the excited state couplings. Modulus of a rephasing 2D spectrum for an inhomogeneous system. (d) Each resonant frequency is decoupled from the others and is represented by a red circle. The homogeneous width of each emitter, represented by the size of the circle, is measurable along the anti-diagonal. (e) The total signal forms an elongated peak along the diagonal (inhomogeneous broadening). (f) Pulse sequence used to perform a MDCS measurement involving 3rd order polarization.}
    \label{fig:2DExampleSpectra}
\end{figure*}
\begin{figure*}[ht]
    \centering
    \includegraphics[width=0.8\textwidth]{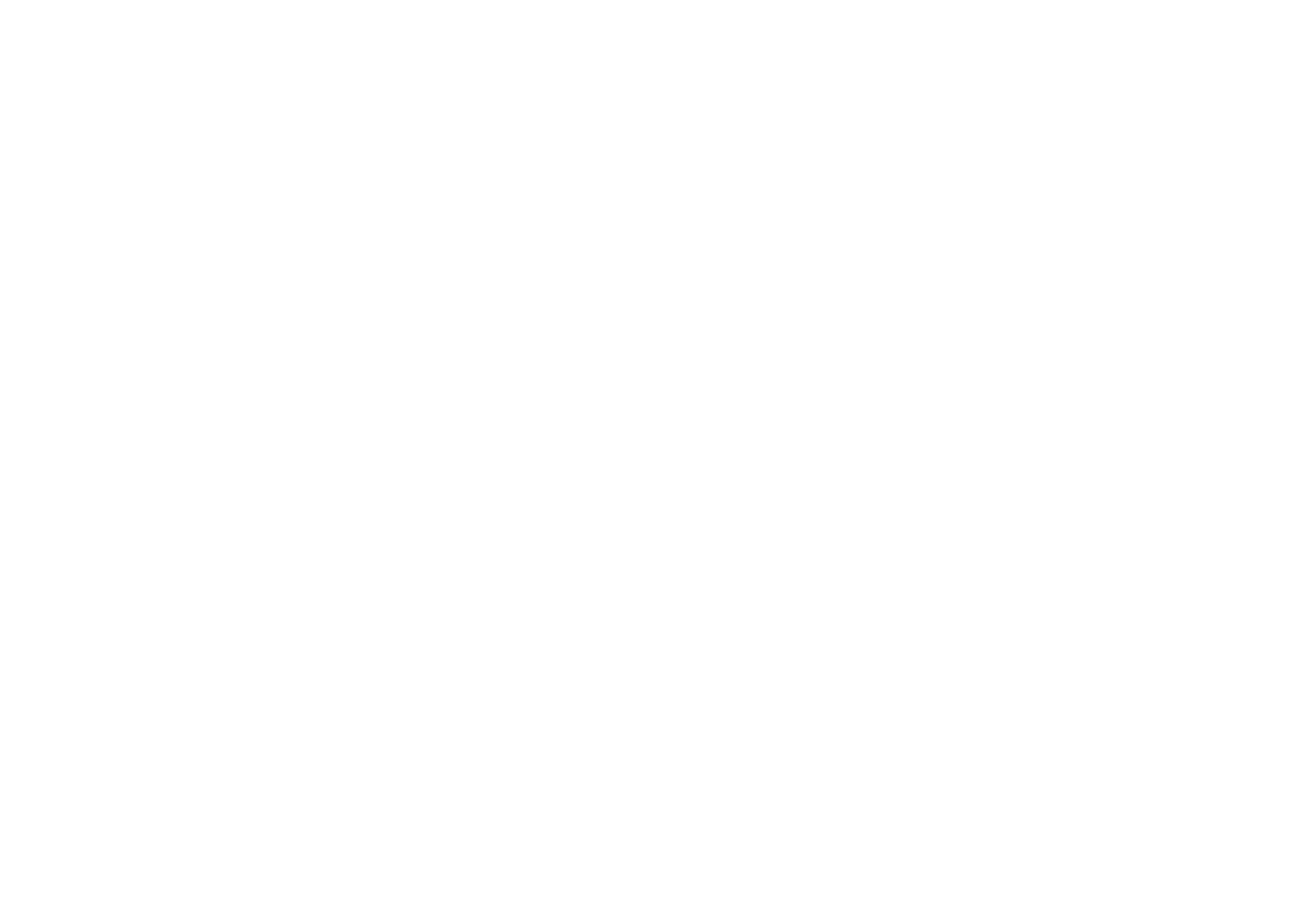}
    \caption{2D amplitude lineshapes for rephasing signals. (a-c) 2D frequency plots for a fixed value of homogeneous broadening with increasing inhomogeneous broadening. The diagonal (red dashes) and cross-diagonal (blue dots) lines are shown. The vertical scale for $\omega_{\tau}$ is negative because of phase-matching requirements, and increases (gets more negative) going down. (d-f) Slices of the corresponding 2D frequency plots along the diagonal (red) and cross-diagonal (blue) directions. The inset compares cross-diagonal slices in the limits of strong homogeneous (dashes) and inhomogeneous (dots) broadening. Figure extracted with permission from Ref.~82.}
    \label{fig:cundiff2010}
\end{figure*}

The optical properties of $\pi$-conjugated polymers are governed by the nature of their primary photoexcitations~\cite{kohler2015electronic}, which are Frenkel excitons~\cite{frenkel1931transformation}. These are strongly bound quasiparticles with binding energies $\gg 10\,k_BT$ at ambient temperature~\cite{pope1984electronic}. Excitonic properties in organic semiconductors in general provide a window into their materials physics, and have always been investigated as a key probe of their electronic landscape~\cite{abe1992singlet,abe1992excitons,yan1994spatially,bredas1996exciton,bassler1997excitons,sariciftci1998primary,rohlfing1999optical,friend1999electroluminescence,schweitzer2000excitons,barford2002theoretical,barbara2005single,scholes2006excitons,bittner2007theory,barford2013excitons,barford2013electronic,Spano2005JChemPhys,spano2006excitons,Spano2006ChPhys,Silva2007PhysRevLett,Silva2009AppPhysLett,spano2009determining,Spano2010AccChemRes,paquin2013PhysRevB,Yamagata2014JPhysChemC,spano2014AnnRevPhysChem,hestand2017molecular,Spano2018ChemRev,ghosh2020excitons,zhong2020unusual,balooch2020vibronic,chang2021hj}. The solid-state absorption spectra of conjugated polymers, while encoding signatures of photophysical aggregates of molecular chromophores along the polymer backbone~\cite{Spano2005JChemPhys,spano2006excitons,Spano2006ChPhys,Silva2007PhysRevLett,Silva2009AppPhysLett,spano2009determining,Spano2010AccChemRes,paquin2013PhysRevB,Yamagata2014JPhysChemC,spano2014AnnRevPhysChem,hestand2017molecular,Spano2018ChemRev,ghosh2020excitons,zhong2020unusual,balooch2020vibronic,chang2021hj}, are often dominated by static disorder, and the large inhomogeneous linewidth obscures details of the rich spectral structure. The seminal work of Spano on excitons in conjugated polymers, of which Refs.~\citenum{Spano2005JChemPhys}--\citenum{chang2021hj} are a subset, involved implementation of Holstein Hamiltonians that treat excitonic coupling, vibronic coupling, energetic disorder, and charge-transfer components when describing push-pull materials, incorporating all of these contributions on equal footing. This has presented a formal way to interpret absorption and photoluminescence lineshapes to quantify excitonic parameters such as free-exciton bandwidth and the nature of their disordered energy landscape, which has been tremendously impactful in semiconductor-polymer science over the past decade. The main observables that enter the linear lineshape interpretation are relative 0--0 and 0--1 peak intensities, which quantify excitonic coupling in the absorption spectrum~\cite{Spano2005JChemPhys,Spano2006ChPhys,Silva2007PhysRevLett,Silva2009AppPhysLett}, with additional information on the correlated nature of the energetic disorder via photoluminescence~\cite{spano2009determining,paquin2013PhysRevB}. To go beyond this effective exciton model, more detailed analysis of the spectral structure is desirable, and techniques such as resonance Raman spectroscopy, for example, are valuable~\cite{tsoi2012effects,martin2015resonance,vezie2016exploring,grey2019resonance}. Many details of the many-body excitonic structure that dominates solid-state behaviour of conjugated polymers are not readily extracted from the broad, convoluted linear optical lineshapes, and it is advantageous to measure nonlinear lineshapes, which resolve rich structure within inhomogeneously broadened spectra beyond linear lineshapes~\cite{Tokmakoff2000,dantus2001coherent,nenov2015spectral,moody2017advances,srimath2020stochastic,srimath2022homogeneous}. Furthermore, many-body excitonic details central to the photophysical aggregate behaviour are not readily identifiable from linear lineshapes; although these states are localized within a chromophore, at sufficiently high densities, exciton-exciton interactions start to dominate the optical properties of organic solids~\cite{agranovich1968collective}. For example, the spontaneous formation of strongly coupled light-matter quantum condensates in organic materials depends fundamentally on the details of exciton-exciton interactions~\cite{keeling2020bose}; how are two-quantum interactions manifested in spectral lineshapes of Frenkel excitons? We have found in early work that both charges and triplet excitons are produced on ultrafast timescales due to multi-quantum interactions in conjugated polymers~\cite{stevens2001exciton,silva2001efficient,silva2002exciton}, but no detail on these can be derived by linear excitation lineshapes. Furthermore, we have found that prompt charge separation occurs via what we have interpreted as charge-transfer states that appear to be intrinsic in these systems~\cite{glowe2010charge,gosselin2011persistent,paquin2011charge,paquin2015multi}, and fundamental electronic couplings that give rise to these is of fundamental relevance. 

Our objective for this article is to discuss two-dimensional coherent spectroscopy as a non-linear optical probe that contributes to the understanding of excitons in $\pi$-conjugated polymer materials, to discuss how this approach is now to be considered a mainstream materials characterization tool in this field, and to review what we consider to be key contributions of this class of techniques to the semiconductor polymer materials science. We first discuss the information that can be discerned from two-dimensional coherent optical lineshapes, and then we review contributions to the understanding of exciton and biexciton structure in neat polymer films, as well as charge separation in heterostructures with electron acceptors designed for photovoltaic applications. We highlight how some photocurrent-detected implementations are particularly valuable in optoelectronic devices. 

\section{Two-Dimensional Coherent Spectroscopy}

%References I used for the sections regarding MDCS and the introduction to non-linear spectroscopy are \cite{Mukamel1995, Turner2014, Lomsadze2020, Nardin2015, Fuller2015, Gregoire2017, li2016probing, Levenson1988, Mueller2020, Tahara2020, Hybl1998, Tokmakoff2000, Bristow2011}.

%\subsection{Linear and Non-linear spectroscopy}
In optical spectroscopy, the interaction between light and matter %can be studied by observing the interaction between an incident electric field and a 
produces an induced mesoscopic polarization $\Vec{P}(t)$ that depends on time according to the electric field component of the driving electromagnetic waves. %The electric field acts on the material dipoles and generates a macroscopic oscillating polarization. 
According to the laws of electrodynamics, this time-varying polarization serves itself as a source of electromagnetic waves, resulting in emission of radiation (measurable signal) with a well-defined wavevector ($\Vec{k}_{S}$) that respects momentum conservation of the incident excitation source, meaning it has a distinct spacial orientation (see Fig.~\ref{fig:colbert} and Ref.~\citenum{thouin2018stable} for an example of this experimental implementation). If the excitation source is coherent, due to laser excitation, this signal has, in principle, a well defined phase relationship with respect to the excitation source, and the dissipation of this coherence limits the optical linewidth. Quantum mechanics establish that the induced polarization is the expectation value of the quantum dipole operator~%following Mukamel's formalism in reference 
\cite{Mukamel1995}, in which the electric fields drive transitions between quantum states of the system. 
It is common to expand the induced polarization in a power series of the incident electric field, taking into account successive light-matter interactions:  
\begin{multline}
    \Vec{P}(t) = \epsilon_0 \biggl ( 
    \chi^{(1)} \Vec{E}_A(t) + \chi^{(2)} \Vec{E}_A(t) \cdot \Vec{E}_B(t)\\ 
    + \chi^{(3)} \Vec{E}_A(t) \cdot \Vec{E}_B(t) \cdot \Vec{E}_C(t) + \ldots 
    \biggr ),
    \label{eq:polarization}
\end{multline}
with the nonlinear susceptibilities $\chi^{(i)}$ (which are tensors), $\epsilon_0$ the vacuum permittivity, and $\Vec{E}_j (t)$ the electric field of the wavepackets of the individual laser pulses that are used to induce the successive light-matter interactions.  
In the first order of this expansion, the polarization is proportional to the incident electric field, which corresponds to the linear response of the sample. All higher orders are considered to be the non-linear response of the material. %In centrosymmetric media such as that of semiconductor polymers, odd-parity nonlinear susceptibilities have non-vanishing amplitude and $\chi^{(3)}$ is the lowest nonlinear response typically probed. 
The most common way to explore the non-linear response is to have multiple electric field interactions with the material. This involves using light pulses with strict time control (inter pulse delay), well-defined phase, and frequency control (pulse spectrum), with a typical experimental geometry as depicted in Fig.~\ref{fig:colbert}. This is one implementation of multi-dimensional coherent spectroscopy (MDCS), but we note that there are a diversity of MDCS; we direct the reader to Ref.~\citenum{Fuller2015} for an excellent overview of the diversity of techniques. The $n$ number of light pulses employed corresponds to the $n^{\mathrm{th}}$ non-linear order, e.g.\ interaction of three light pulses corresponds to the $3^{\mathrm{rd}}$ order non-linear response of the material, five light-matter interactions correspond to the $5^{\mathrm{th}}$ order non-linear response, and so on. The work on conjugated polymers has been limited to probing the third order response, and we thus limit our discussion on this order of nonlinear spectroscopy.

%\subsection{Overview of multidimensional coherent spectroscopy}

In MDCS, the non-linear signal contains information regarding ultrafast dynamics of electronic and vibrational phenomena of a system~\cite{li2016probing, Gregoire2017, Lomsadze2020, Turner2014, Mueller2020}. Some advantages over linear spectroscopy techniques are the ability to resolve homogeneous and inhomogeneous broadening of the spectral features independently from each other, the separation of interaction pathways as well as the identification of correlations (coupling) between exited states of the system~\cite{Mukamel1995, Nardin2015, Fuller2015, Tahara2020, Hybl1998, Tokmakoff2000, Bristow2011}. Additionally, the signal detection is not limited only to coherent light~\cite{Levenson1988}, but in can also be implemented in order to detect incoherent emission~\cite{Tekavec2007JCP, Cundiff2013, Widom2013}, photocurrent~\cite{Cundiff2013,Karki2014,Bakulin2016,vella2016ASciRep,li2016probing,Gregoire2017,Bargigia_GaAs}, steady-state photoinduced absorption~\cite{li2016probing} or transient absorption~\cite{Cerullo2011}.

For the case of isotropic media the second order MDCS signal vanishes, consequently the most common measurement is of third order susceptibility $\chi^{(3)}$ in equation~\ref{eq:polarization}~\cite{Mukamel1995}. Three light pulses are required in order to obtain a third order non-linear response, with geometry as depicted in Fig.~\ref{fig:colbert}(a), and again in Fig.~\ref{fig:2DExampleSpectra}(f). The first pulse, creates a coherent superposition between the ground and an excited state of the system, also known as coherence. The phase of this coherence oscillates at the frequency difference between the ground and excited states during a time ($\tau_{1}$), and mesoscopically, a coherent polarization is generated. The second pulse creates a population state that evolves during a time $\tau_{2}$. Depending on the experiment, instead of population, the second pulse can create further coherences between higher excited states. Finally, the third pulse creates a new coherence between ground and excited states, which radiates the measured signal after a time $\tau_{3}$ \cite{Turner2014}. This is represented in Figs.~\ref{fig:colbert}(b), \ref{fig:colbert}(c), and \ref{fig:2DExampleSpectra}(f). 

A MDCS experiment records the phase oscillations of the coherences created by pulse 1 and measured with pulse 2, as a function of the delay time $\tau_{1}$. Subsequently, a 2D spectrum is extracted using a %multidimensional 
Fourier transform along that time variable. The axis related to the Fourier domain of $\tau_{1}$ %and $\tau_{3}$ are 
is referred to as \emph{excitation} (or \emph{absorption}) %and emission 
energy. %, respectively. 
The spectral amplitude and phase of the coherent signal emitted by the nonlinear material response is measured with a spectrometer and heterodyne detection (spectral interferometry) with the LO pulse, and this spectral axis is referred to as the \emph{emission} energy, and these spectral interferograms are measured as a function of $\tau_1$, usually at fixed $\tau_2$, but alternatively in the case of two-quantum coherence measurements, at fixed $\tau_1$ and as a function of $\tau_2$ (Fig.~\ref{fig:colbert}(a)). 
Fig.~\ref{fig:2DExampleSpectra} shows a simulation of typical MDCS 2D maps considering a system of two excited states. The features along the diagonal correspond to optical transition autocorrelations, which reproduce the structure of a common linear absorption spectrum \cite{Bristow2011}. 
Off diagonal features (cross-peaks) are created when different excited states share a common ground state (Fig. \ref{fig:2DExampleSpectra}(a)), meaning that it is an intrinsic feature of the material and not an indication of sample inhomogeneity. The absence of cross peaks (Fig.~\ref{fig:2DExampleSpectra}(b)) reflects two uncorrelated optical transitions. Variations in the intensity and shape of these cross-peaks provides information on the coupling (correlation) strength between the exited states (Fig. \ref{fig:2DExampleSpectra}(c)).

Measurements can be classified as N-quantum depending on the difference in the number of quanta separating the two states involved in the generated coherences. The most common example is a one-quantum (1Q) measurement where the ground and excited state are separated by one quantum of the incident photon energy. In this case the MDCS 2D map probes directly the optical transitions related to absorption and photoluminescence phenomena of the sample \cite{Mueller2020}.

%However, t
To probe excited states lying two-quantum above the ground state arising from e.g.\ biexcitons, it is necessary to perform two-quantum (2Q) measurements. Depending on the MDCS experiment, when  having a non-collinear beam configuration, is possible to isolate two-quantum coherent signals by using a specific pulse sequence as depicted in Fig.~\ref{fig:colbert}(c)~\cite{stone2009exciton,moody2014coherent,elkins2017biexciton,thouin2018stable}. While, for a collinear beam configuration this can be achieved by doing a second harmonic lock-in demodulation~\cite{bruder2015efficient,gutierrez2021exciton,meza2021molecular}, this experiment has demonstrated to be sufficiently sensitive to probe biexciton states in a polymeric system~\cite{meza2021molecular,gutierrez2021exciton}.

The 2D spectra measured with MDCS provides us with the ability to separate the homogeneous and inhomogeneous broadening contributions of the spectral linewidth, in contrast to linear spectroscopy techniques, where the spectral features are a combination of both types of broadening~\cite{Mukamel1995,yajima1979JPSJ}. The homogeneous linewidth is directly related to the time decay of the coherence generated by the light-matter interaction (also known as dephasing time). Meanwhile, in the case of inhomogeneous broadening, the linewidth is a result of the contributions from the distribution of resonant frequencies, brought on by sample inhomogeneities, defects, Doppler broadening, etc.\cite{Levenson1988,Fuller2015,Cundiff2010OptExp}

Figs.~\ref{fig:2DExampleSpectra}(d), (e) show an example of a 2D spectra where a distribution of decoupled emitters, each one with its own homogeneous width (red circles), leads to the emergence of a spectral feature with inhomogeneous (homogeneous) broadening corresponding to the diagonal (cross-diagonal) width of the measured peak, in the limit of strong inhomogeneity. 
Typically, the way in which the inhomogeneous (homogeneous) linewidths are obtained is by measuring the peak width along a cut in the diagonal (cross-diagonal) of the 2D spectra. However, this procedure is not entirely accurate in the limit in which inhomogeneous broadening is not the dominant contribution. As Siemens et~al. showed~\cite{Cundiff2010OptExp}, there is coupling between the diagonal and cross-diagonal spectral lineshapes, i.e.\ an increase of the inhomogeneous broadening will also lead to a slight elongation in the cross-diagonal direction, and vice versa. The authors demonstrate a method, using projection-slice theorem, to extract the absolute homogeneous and inhomogeneous linewidth of the spectra. They considered a Markovian approximation~\cite{Mukamel1995} for the homogeneous broadening (resulting in an exponential decay with a dephasing rate $\gamma$), and a Gaussian envelope of with $\sigma$ for the inhomogeneous broadening. 
The authors showed that in the case where the distribution of resonant frequencies is much larger than the dephasing rate ($\sigma \gg \gamma$), the system is in an inhomogeneous limit. The amplitude of a diagonal cut follows a Gaussian form of width $\sigma$, whilst a cross-diagonal cut has a square root of a Lorentzian shape with a width $2\gamma$. This is shown in Fig.~\ref{fig:cundiff2010}(c) and (f), respectively. In this limit, the linewidths can be extracted with minimal error using the typical method previously mentioned. On the other hand, in a purely homogeneous limit ($\gamma \gg \sigma$), the authors showed that both the diagonal and cross-diagonal cuts have the same Lorentzian shape (width $2\gamma$), as shown in Fig.~\ref{fig:cundiff2010}(a) an (d). Finally, in the special case where the homogeneous and inhomogeneous broadening are similar ($\sigma \sim \gamma$), the diagonal and cross-diagonal amplitudes follow a more complicated shape, as shown in Fig.~\ref{fig:cundiff2010}(b) and (e). The diagonal amplitude is a convolution of a Gaussian and a Lorentzian function, whilst the cross-diagonal amplitude takes the shape of a complementary error function. 

Conjugated polymers referred to in this work are typically in a inhomogeneous broadening limit~\cite{Pascal2017PRB,song2015JCP}. This means that the intrinsic homogeneous (inhomogeneous) broadening can be obtained by performing a zero-time rephasing experiment, and measuring the Lorentzian (Gaussian) width of the cross-diagonal (diagonal) feature. In this type of measurement, the phase evolution of the coherence during $\tau_{1}$ and $\tau_{3}$ are opposite, so any non-intrinsic inhomogeneous dephasing that occurs during $\tau_{1}$ is cancelled by a mirror-opposite rephasing during time $\tau_{3}$. If the population time is set as $\tau_{2}=0$, then the measurement is referred to as photon-echo. We will consider below an example of a 2D coherent spectral lineshape that is \emph{not} indicative of dominant inhomogeneous broadening, although this would not have been evident from linear spectral lineshape analysis. 

We highlight that in spite of efforts by some members of the MDCS community to establish conventions, there are currently none for displaying MDCS plots, as different authors display such spectra in different ways. (Even the current authors have adopted different conventions over time.) The published data displayed in this review reflect that sociology. 

\section{Excitons in neat polymers}\label{sec:excitons}

\subsection{Linear optical signatures of excitons in photophysical aggregates}

\begin{figure}[htp]
    \centering
    \includegraphics[width=\columnwidth]{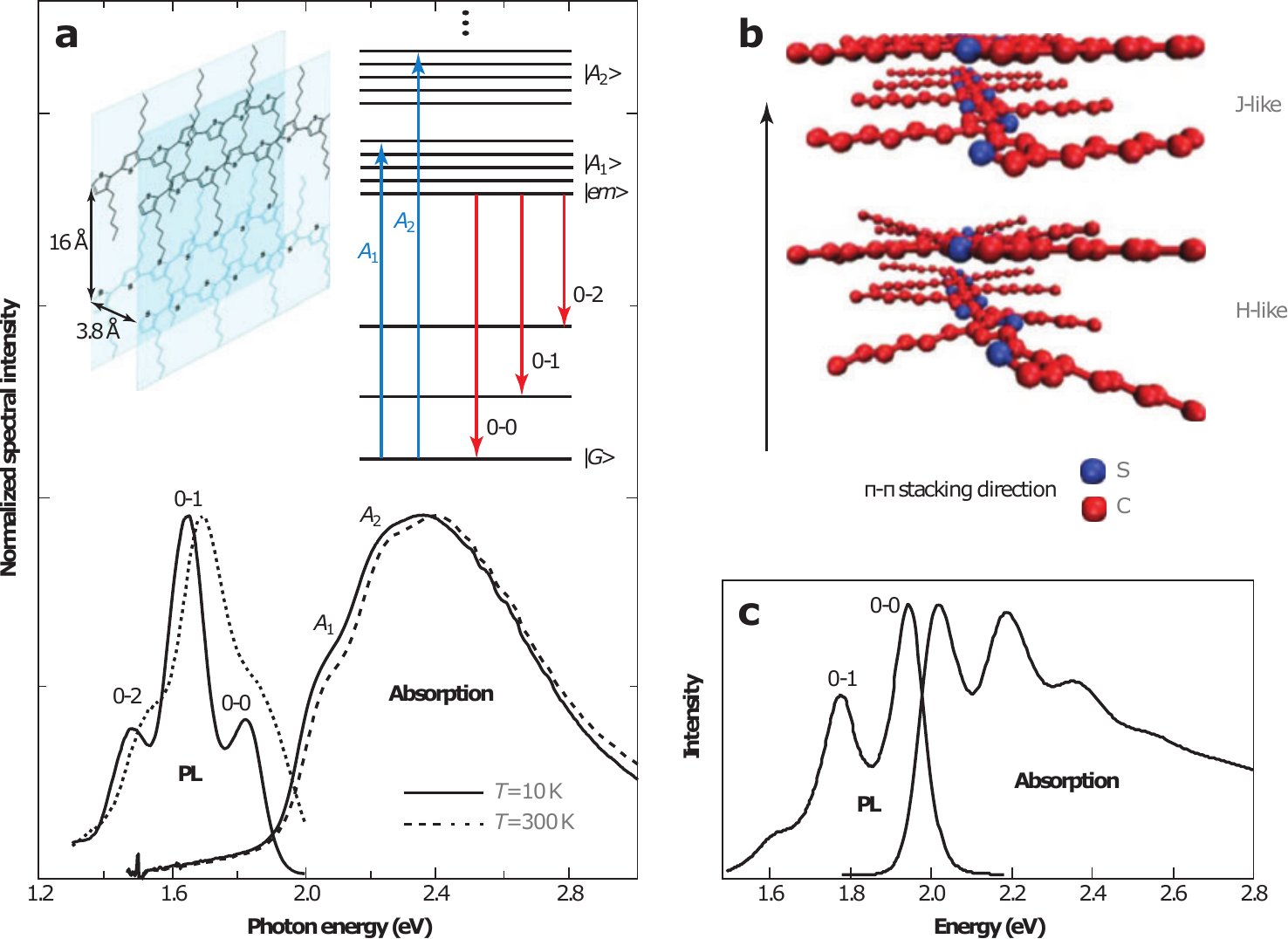}
    \caption{(a) Absorption and photoluminescence (PL) spectra of a P3HT film cast from a chloroform solution. %Panel a reprinted with permission from Ref. \cite{spano2009JChemPhys}. 
    (Insets) P3HT $\pi$-stack and Jablonski diagram corresponding to the weakly coupled H-aggregate model. (b) Graphical depiction of P3HT J- and H- aggregates. (c) Absorption and PL spectra of P3HT nanofibers grown in toluene. %Panel c reprinted with permision from Reference~\cite{niles2012JPhysChemLett}. 
    The Figure was extracted with permission from Ref.~\citenum{spano2014AnnRevPhysChem}.}
    \label{fig:Hagg}
\end{figure}

The photophysics of conjugated polymer assemblies can be understood in the context of Kasha theory \cite{kasha1965exciton,spano2014AnnRevPhysChem}. In this model Kasha et~al.\ describe the role of optically generated excitations in the absorption and fluorescence of small chromophore aggregates. Briefly, Coulombic coupling is mediated via dipole-dipole interactions between polymer chromophores, leading to the formation of excitonic states. If the coupling term is negative, the neighbouring polymer chains have a head-to-tail orientation and are referred to as J-aggregates. On the other hand, if the coupling term is positive, the chain arrangement is side-by-side and the polymer is known as H-aggregate. 

Excitons have different physical traits depending on the geometric arrangement of the chains \cite{Spano2018ChemRev,spano2014AnnRevPhysChem}. In J-aggergates (along the chromophore chain), excitons are of Wannier-Mott characteristics, meaning they can be separated over several repeated units. For H-aggregates (across the chains), the excitons have more Frenkel-like properties, where the separation is limited to neighbouring chains \cite{Yan1994PhysRevLett, Yan1992PhysRevLett, Spano2010AccChemRes}.

Extensions to the Kasha model include the effect of intramolecular vibrations on the spectra of exciton states \cite{Spano2005JChemPhys}. These vibrations are often referred to as progression-building due to the characteristic succession of peaks that appear in the absorption/emission spectra, as shown in Fig.~\ref{fig:Hagg} for poly(3-hexylthiophene) (P3HT) films. Because of this, P3HT has been used as model platform to develop the physical understanding of vibrational coherence and electron transfer dynamics in organic photovoltaics~\cite{song2015JCP}. Its semi crystalline nature gives rise to the previous mentioned vibrational structure, and its strong coupling with optical excitons are visible in the absorption/emission spectra~\cite{song2015JCP}.

P3HT aggregates formed from good solvents are mostly H-like (see Fig.~\ref{fig:Hagg}), favoring interchain interactions. This is likely due to the abundance of aggregates with low conjugation lengths, consequence of a high level of disorder brought on by fabrication processes~\cite{Spano2005JChemPhys, Silva2007PhysRevLett, Silva2009AppPhysLett} (rapid solvent evaporation, spin-casting films, etc) \cite{spano2014AnnRevPhysChem}.

The underlying differences between J- and H-aggregates are reflected in the features of the absorption and photoluminescence (PL) spectra of the polymers. The clearest example is the Frank-Condon vibronic progression resulting from coupling between the vinyl-stretching modes of the $\pi$-conjugated backbone and the optical exciton $S_{0} \rightarrow S_{1}$ transition~\cite{Spano2018ChemRev}. Understanding this vibrational structure of the absorption and PL spectra has been a key approach to explain structure-property relations. Any distortion on the vibronic progression upon aggregation helps predict the effect on the spectral features due to disorder, temperature and molecular assembling~\cite{Nelson2013JPhysChemB,Knoester1993JChemPhys, Meskers2000ChemPhys,Cerullo2017JPhysChemLett,Spano2014JPhysChemLett,Spano2011JPhysChemB}.

%\subsubsection{Photoluminescence}
Regarding the PL spectrum, it is considered that the emission occurs from the lowest energy exciton following Kashas rule~\cite{kasha1950}. In the absence of disorder, this corresponds to an exciton with wavevector equal to zero and vibrational quanta $\nu=0$ (labeled as 0--0 in Figure~\ref{fig:Hagg}). Subsequent spectral peaks are referenced according to the total number ($\nu$) of vibrational quanta. The relative intensity of the 0-0 peak with respect to the rest of the vibronic progression is directly related to the exciton coherence length (distance over which an exciton maintains wavelike behaviour)~\cite{Scholes2011ChemMat, Panitchayangkoon2010ProceedNAS,paquin2013PhysRevB}.

Specifically in J-aggregates, higher exciton coherence lengths lead to constructive interference between the emitting dipoles resulting in an enhancement of the 0-0 peak, whilst for H-aggregates, the interference is destructive, causing a suppression of the 0-0 spectra peak~\cite{spano2014AnnRevPhysChem}. Therefore, an ideal H-aggregate has a PL spectrum formed by a vibronic progression without the 0-0 peak, whereas the PL spectrum of a J-aggregate is also a modified Frank-Condon progression, dominated by a superradiant 0-0 peak~\cite{Silva2007PhysRevLett, De1990ChemPhysLett,fidder1990ChemPhysLett}.

The PL spectra is also sensitive to the amount of disorder in the conjugate. In P3HT, this disorder comes in the form of structural (packing) defects, chemical impurities (oxidized defects), torsional defects~\cite{Feng2007PolyEngSci, Salaneck1988JChemPhys,Lucke2015JPhysChemB,Mullerova2016SolEn}. In H-aggregates, where the 0-0 peak is suppressed, the introduction of disorder allows for 0-0 emission. Meaning that the strength ratio between 0-0 and 0-1 emission is direct probe for disorder~\cite{Spano2009JCP}.

%\subsubsection{Absorption}

Another defining difference between J- and H- aggregates is observable in the absorption spectrum. The ratio of the $A_{1}$ and $A_{2}$ absorption peaks is related to the nearest-neighbor interchain Coulomb coupling, which is directly proportional to the exciton bandwidth~\cite{Silva2009AppPhysLett}. In the case of J-aggregates, an increasing ratio reflects an increase in the exciton bandwidth as well as a redshift of the main peak ($A_{1}$). For H-aggregates, the behaviour is opposite, i. e., the increase ratio of the $A_{1}$ and $A_{2}$ correlates to a decrease in the exciton bandwidth and a blue-shift of the spectrum~\cite{Spano2006ChPhys,Spano2010AccChemRes}.

Previous studies have found that the exciton bandwidth depends on the interchain Coulomb coupling, which at the same time depends on the conjugate chain length $L$. Once $L$ exceeds the intermolecular separation, the interchain coupling goes down as a function of $L$, i. e. shorter chains lead to stronger interchain interaction and more H- like behaviour~\cite{ Scharsich2012,Gierschner2009,Barford2007,Beljonne2000,Spano2009JCP,Silva2009AppPhysLett}.

In an ideal case where the chains are infinitely long, the interchain interaction vanishes. This indicates that the absorption spectra can directly probe the interchain coupling and intrachain order. 
In practice, one has to take into account concurrently inter- and intra-chain interactions in the calculations. Therefore a realistic picture of a polymer aggregate is one that has both a J- and H- characteristics, meaning the polymer is a 2D excitonic system containing delocalized excitons along the polymer chain as well as between the chains.
Despite the valuable information that can be extracted from absorption and photoluminescence spectra, the details on how excitons interact with their environment cannot be obtained from the linear spectra. This is mainly because the homogeneous and inhomogeneous line broadening cannot be selectively extracted from the total spectral width. In this context, MDCS has been an important tool to disentangle these broadening contributions. 

In the following sections we will discuss results regarding the photoexcited dynamics of neat polythiopehenes, and their blends, as well as population effects on the spectral linewidths probed via MDCS experiments. 

%\section{2D spectroscopy results on P3HT films}

\subsection{2D coherent excitation lineshape of HJ aggregates}\label{sec:PBTTT_lineshape}

\begin{figure}
    \centering
    \includegraphics[width=0.85\columnwidth]{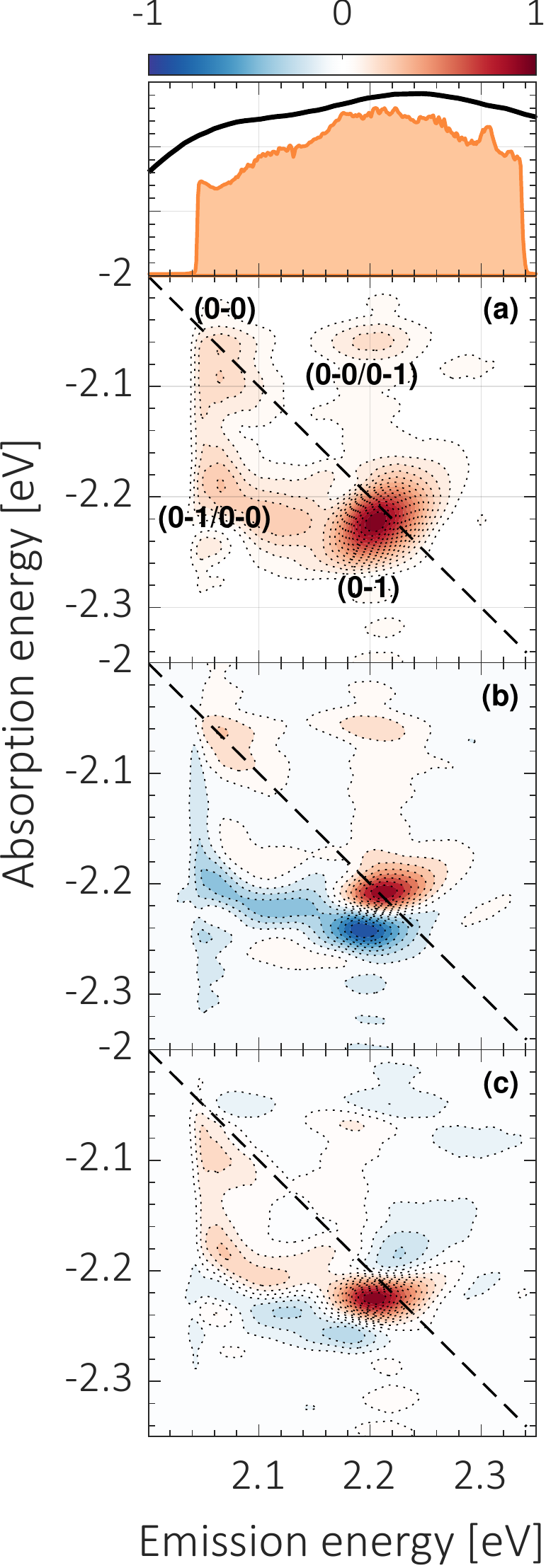}
    \caption{(a) Absolute rephasing 2D coherent spectrum of PBTTT at zero-population-waiting time, measured at 5\,K. The 0--0 and 0--1 diagonal peaks, as well as their cross peaks, are indicated in the figure. (b) Real and (c) imaginary components of the spectrum in part (a). Top: absorption spectrum of PBTTT and spectrum of the femtosecond pulses used in the measurement. Previously unpublished data. }
    \label{fig:PBTTT_lineshape}
\end{figure}

We start our discussion of spectral lineshapes by considering MDCS measurements on a relatively rigid polythiophene derivative.   
The 2D coherent rephasing spectrum of [poly(2,5-bis(3-hexadecylthiophene-2-yl)thieno[3,2-b]thiophene)]~\cite{mcculloch2006liquid}, a conjugated polymer in the polythiophene family, at zero population-waiting-time, i.e.\ $\tau_2 = 0$ in Fig.~\ref{fig:2DExampleSpectra}(f), is displayed in Fig.~\ref{fig:PBTTT_lineshape}. Part (a) displays the norm of the spectrum, while parts (b) and (c) display the real and imaginary spectra, respectively. In this sample, the linear lineshape (top panel) is very similar to that of P3HT in Fig.~\ref{fig:Hagg}(a), and is indicative of dominant H-aggregate behaviour~\cite{spano2014AnnRevPhysChem}. The diagonal 0--0 feature in Fig.~\ref{fig:PBTTT_lineshape}(a) is suppressed with respect that corresponding to 0--1, consistent with the lineal lineshape. Cross peaks between the 0--0 and 0--1 diagonal peaks are also observed, as expected for a vibronic progression, which is the situation depicted in Fig.~\ref{fig:2DExampleSpectra}(a). The real and imaginary spectra, however, show structure in the (0--1/0--0) cross peak, with clear evidence of an excited-state absorption superimposed with the cross peak, a feature not observed in the (0--0/0--1) cross peak. This excited-state absorption is associated with two-quantum transitions, which will be discussed in Section~\ref{sec:biexcitons} below. Furthermore, the norm of the spectrum, shown in Fig.~\ref{fig:PBTTT_lineshape}(a), displays highly symmetric lineshape of the 0--0 and 0--1 diagonal peaks, which is indicative of moderate inhomogeneity, a regime depicted in Figs.~\ref{fig:cundiff2010}(b) and diagonal and anti-diagonal cuts in Fig.~\ref{fig:cundiff2010}(e). From this lineshape, the homogeneous and inhomogeneous linewidths have been extracted, and it was concluded that the importance of the homogeneous contribution to the total linewidth is indicative of the importance of dynamic disorder in this polymer system~\cite{gutierrez2021exciton}. The nonlinear coherent lineshape thus reveals richer information than available via the linear spectrum.

\subsection{Homogeneous dephasing and exciton delocalization}

\begin{figure}
    \centering
    \includegraphics[width=0.8\columnwidth]{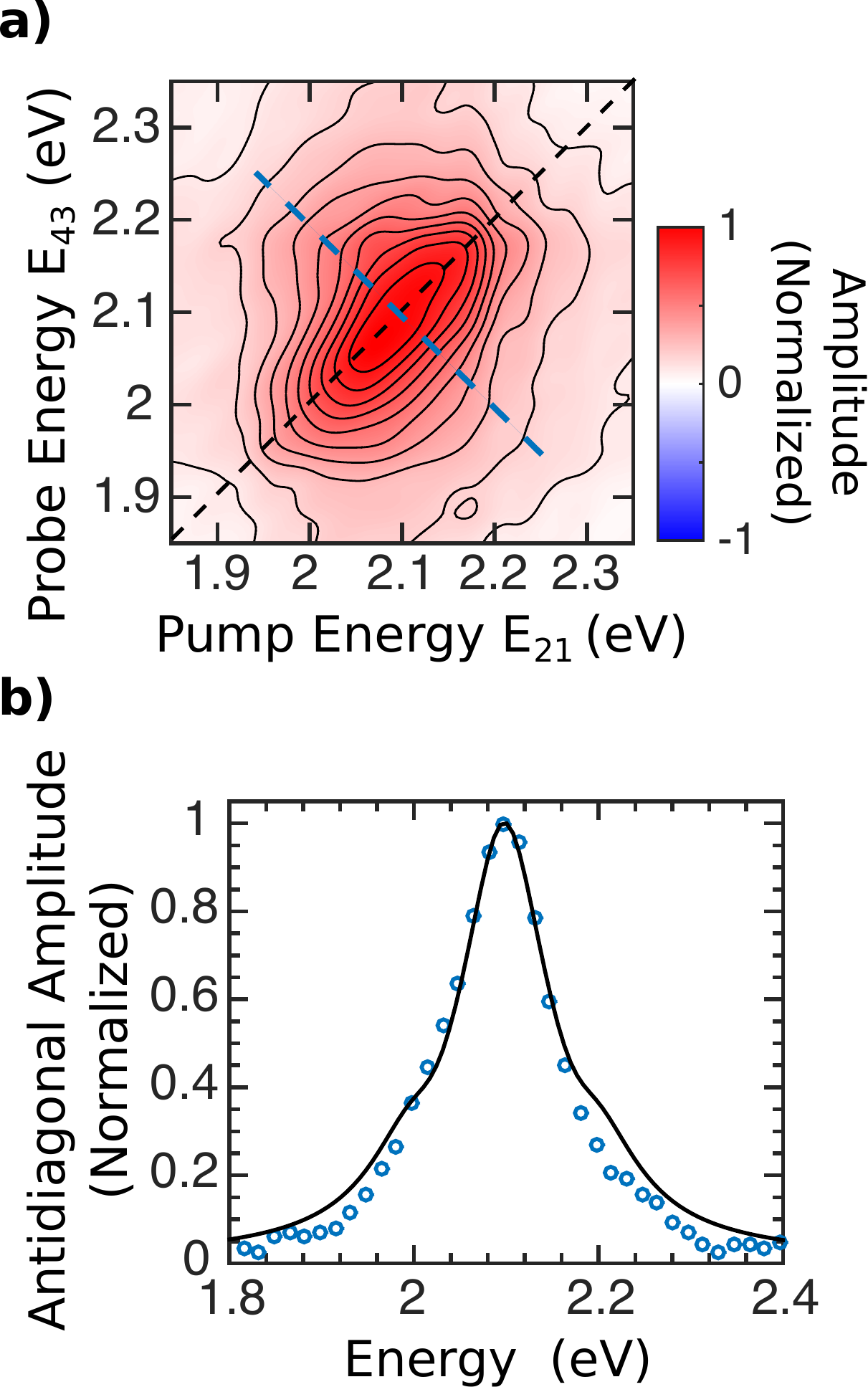}
    \caption{(a) Modulus of the measured 2D-PLE rephasing spectrum. The elongated feature along the diagonal is characteristic of inhomogeneous broadening, limited by the excitation laser spectrum, while the antidiagonal slice (blue dashed line) reveals the homogeneously-broadened spectrum.  (b) Antidiagonal slice of the photon echo signal at 2.10\,eV (blue open circles) corresponding to the dashed blue line in part (a). The black continuous curve displays the corresponding slice for the simulated spectra using $2\gamma=90 \,\text{meV}$ and $\sigma=130$\,meV (shown in Fig~\ref{fig:dephasing_sim}(c)). Figure extracted with permission from Ref.~\citenum{Pascal2017PRB}}
    \label{fig:P3HT_homo}
\end{figure}
\begin{figure}
    \centering
    \includegraphics[width=0.8\columnwidth]{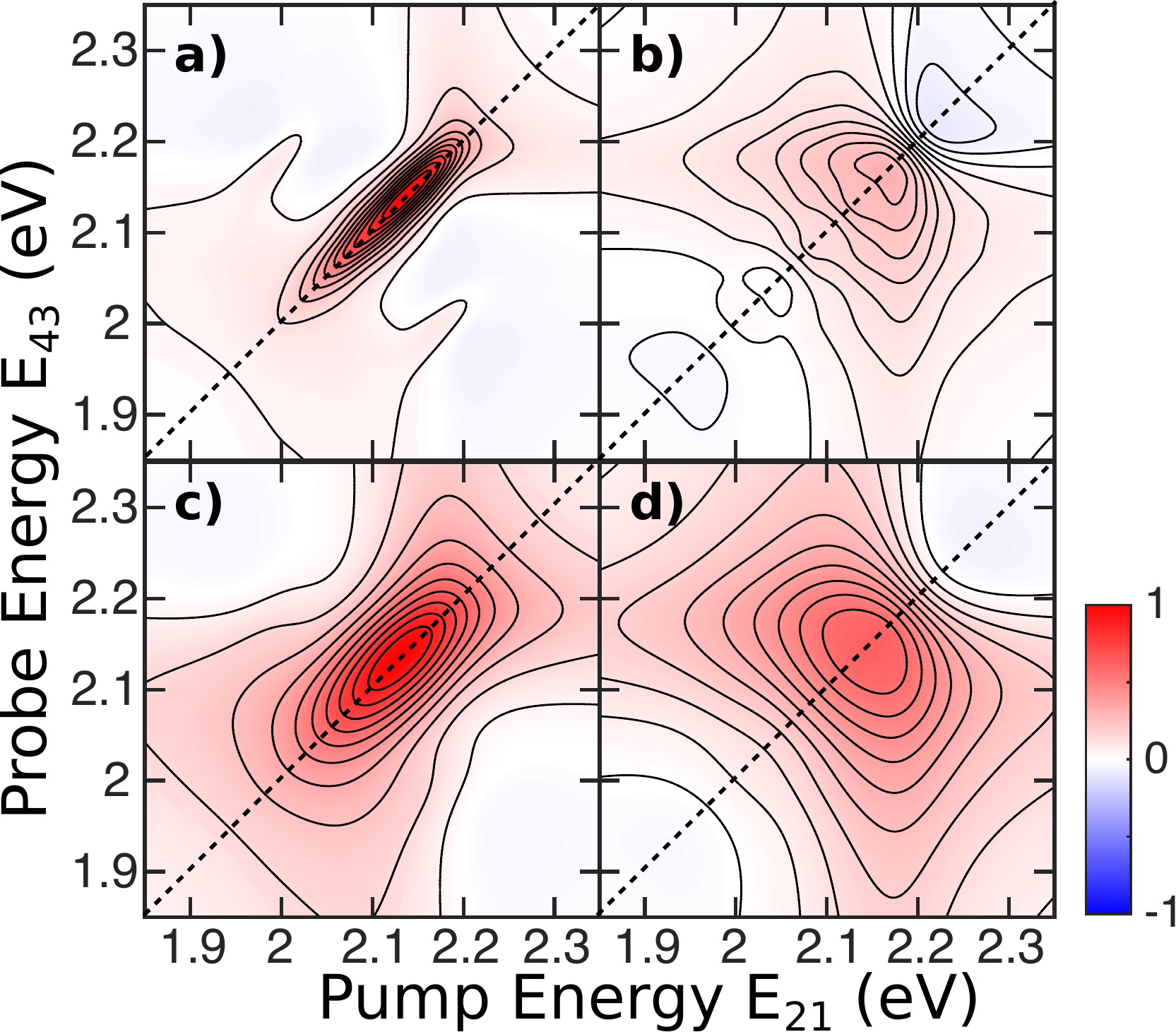}
    \caption{Numerical simulations of the P3HT rephasing (left) and nonrephasing response (right). The real part is displayed using  $2\gamma=30$\,meV (top) and $2\gamma=90$\,meV (bottom), both with inhomogeneous broadening $\sigma=130$\,meV. Here $\gamma$ is the dephasing rate and $\sigma$ is the inhomogeneous spectral width. Figure extracted with permission from Ref.~\citenum{Pascal2017PRB}}
    \label{fig:dephasing_sim}
\end{figure}

%Excitonic models based on photophysical aggregates, have been able to succesfully describe the spectral absorption and photoluminescence lineshapes, however single-molecule spectroscopy is significantly different. 
Single-molecule spectra of conjugated aggregates differs appreciably from its solid state counterparts, in such manner that the spectral lineshape is not due solely to a distribution of inhomogeneous resonators \cite{Thiessen2013,steiner2014PhysRevLett,cook2008EESci}. Previous studies on isolated P3HT chains at low temperature (4\,K) showed that the PL spectrum spans over a large spectral range in the visible region, with a linewidth ranging from $\sim$10-30\,meV and almost invariant line shapes~\cite{Thiessen2013}. Thiessen et~al.~\cite{Thiessen2013} suggested that the inhomogeneous broadening of the PL emission of solution P3HT arises from the conformational disorder at the single chain level. They hypothesized that intrachain torsional disorder gives rise to a distribution of chromophore energies through exciton delocalization and electron-hole polarization~\cite{Simine2017JPCLett}. Consequently, this energetic disorder is responsible for the variability in the spectral range of the emission spectra, leading to a large inhomogeneous linewidth~\cite{Thiessen2013}. This is in contrast to bulk films, where the line shapes are very strongly dependent on the film microstructure~\cite{paquin2013PhysRevB}. The authors of Ref.~\citenum{Thiessen2013} argued that the emission spectral lineshape in the bulk are composed of a distribution of single-chain emission spectra that are distributed throughout the entire visible range, and \emph{not} dictated by the aggregate mode of Spano~\cite{Spano2005JChemPhys,spano2006excitons,Spano2006ChPhys,Silva2007PhysRevLett,Silva2009AppPhysLett,spano2009determining,Spano2010AccChemRes,paquin2013PhysRevB,Yamagata2014JPhysChemC,spano2014AnnRevPhysChem,hestand2017molecular,Spano2018ChemRev,ghosh2020excitons,zhong2020unusual,balooch2020vibronic,chang2021hj}. That report highlighted the need to experimentally isolate the homogeneous spectral linewidth of P3HT in the solid state. 

In a subsequent study, Gregoire et~al.~\cite{Pascal2017PRB} measured simultaneously the homo/inhomogeneous excitation linewidth of a P3HT bulk film and compared them to those obtained by single-molecule measurement, in order to determine the validity of aggregate excitonic models and the effects of aggregation. They employed an MDCS variant referred to as two dimensional photoluminescence (2DPL), where the measured nonlinear response is the time-integrated PL intensity. The authors demonstrated that the homogeneous linewidth was an order of magnitude larger compared to single polymer chains (at 8\,K). This can be seen in Fig.~\ref{fig:P3HT_homo}(a), which presents the modulus of a 2D rephasing spectra. % displays the antidiagonal (dashed blue line) and diagonal (black dashed line) spectral slices that correspond to the homogeneous and inhomogeneous linewidths, respectively. 
Fig.~\ref{fig:P3HT_homo}(b) shows the extracted homogeneous linewidth (blue dots) compared with simulations performed with a homogeneous dephasing rate of $2\gamma=90$\,meV (black line). The obtained dephasing rate is close to an order of magnitude larger than the lower limit reported for isolated chains~\cite{Thiessen2013}. This establishes unabmiguously that the bulk spectral lineshape is \emph{not} due to a distribution of non-interacting single-chain spectra, but by solid-state, aggregate dispersion that results in more rapid homogeneous dephasing~\cite{Pascal2017PRB}. 

In addition, Gregoire~et~al.~\cite{Pascal2017PRB} reported numerical simulations of the 2D rephasing and nonrephasing spectra, both with homogeneous dephasing rates of $2\gamma=30$\,meV and $2\gamma=90$\,meV (see Fig~\ref{fig:dephasing_sim}). The authors %made a spectra comparison, showing 
demonstrated that a homogeneous linewidth of $2\gamma=30$\,meV, which would correspond to an upper limit of the single-chain linewidth~\cite{Thiessen2013}, leads to a significantly narrower 2D spectra compared to the measurements. %, even though it corresponds to the upper limit reported by Thiessen~et~al.~\cite{Thiessen2013} for single chains. 
The simulations %both 
considered a inhomogeneous linewidth of $\sigma=130$\,meV which corresponds to the disorder widths previously used to model absorption and PL lineshapes in P3HT~\cite{paquin2013PhysRevB}. The authors suggested that the additional homogeneous broadening seen in the bulk film comes from interchain photophysical aggregate effects, which consists of fluctuations in excitonic coupling between $\pi$ electrons across chains \cite{Yamagata2014JPhysChemC}, increasing the dephasing rate and promoting H-aggregate-like signatures in the spectra.

%\textcolor{red}{Expand the discussion to include Reid Chem of Matt paper, Cerullo paper, Son Jurnal of Chemical paper.}

%Examples of how this shows up in P3HT as in Fig.~\ref{fig:song_P3HT} [Song, Nature Communications https://doi.org/10.1038/ncomms5933; Reid Chemistry of Materials https://doi.org/10.1021/cm4027144, cerullo paper; Song Journal of Chemical https://doi.org/10.1063/1.4916325].

\subsection{Exciton population dynamics in P3HT}

\begin{figure}[h]
    \centering
    \includegraphics[width=\columnwidth]{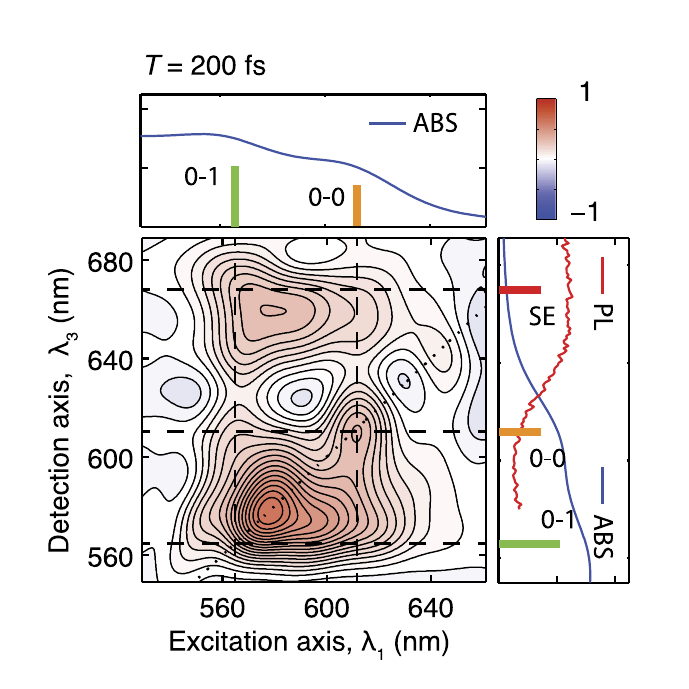}
    \caption{Representative absorptive 2D spectra of P3HT film with a population time $\tau_2=200$ fs, along with the linear absorption and photoluminescence spectra to illustrate the peak positions. Figure extracted from Ref. \cite{song2015JCP}}
    \label{fig:song_P3HT}
\end{figure}

\begin{figure*}
    \centering
    \includegraphics[width=0.80\textwidth]{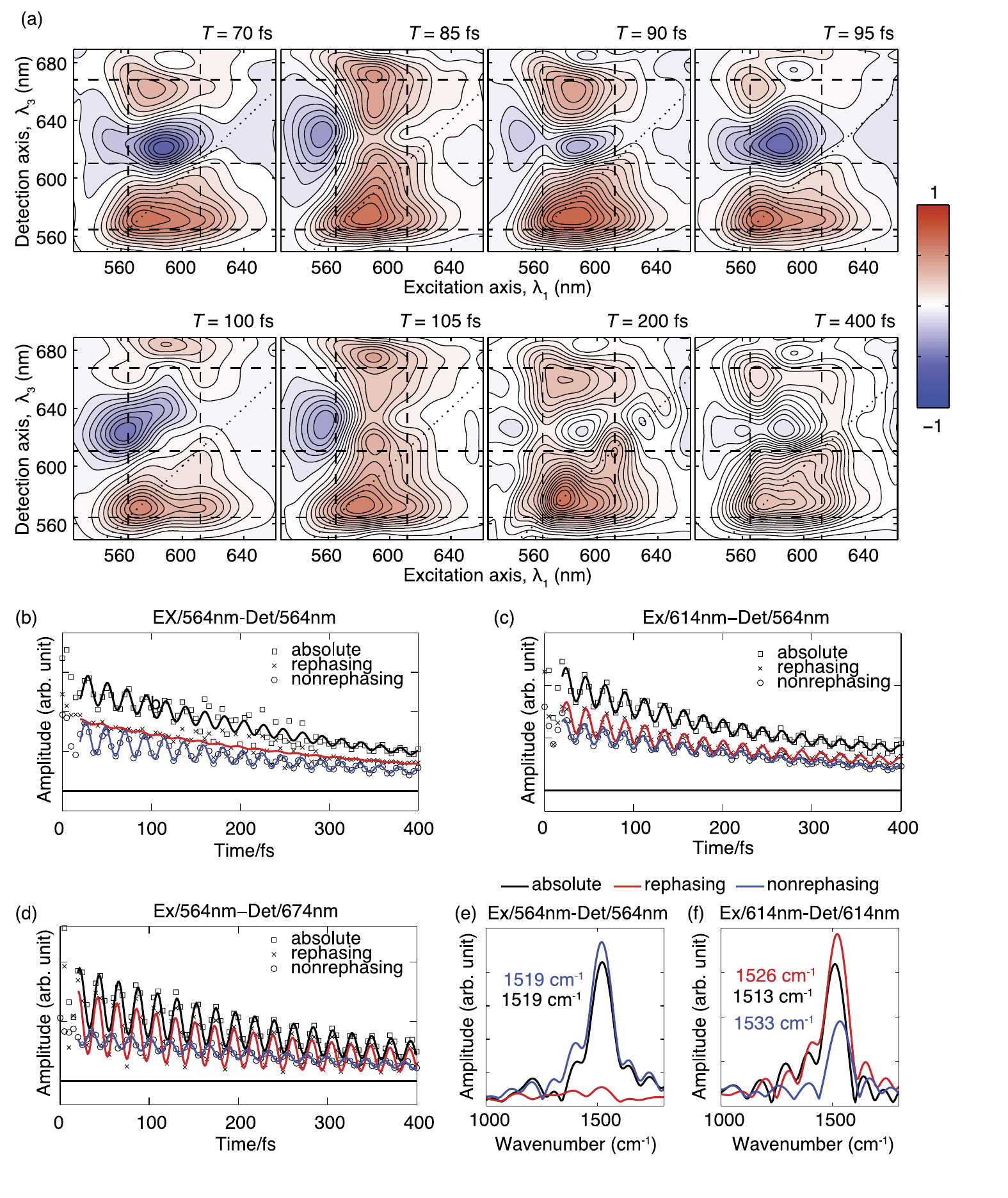}
    \caption{(a) Real part of the MDCS 2D maps of a P3HT film at different population times $\tau_2$. (b)-(d) Time traces of the amplitudes of one diagonal peak (with excitation and detection wavelengths at 564 nm) and two cross peaks (with excitation wavelength at 614 nm (or 564 nm) and detection wavelength at 564 nm (or 674 nm)). Time traces from the rephasing spectra (absolute value, cross), nonrephasing spectra (absolute value, empty circle), and the 2D spectra (absolute value, empty square) are shown. (e) and (f) Correspond to the Fourier transforms of the time traces of two diagonal peaks (with excitation and detection wavelengths at 564 nm (or 614 nm)) along the population time axis. Figure extracted from Ref.~\citenum{song2015JCP}. }
    \label{fig:scholes}
\end{figure*}

MDCS has been shown to be a powerful method to study the sub picosecond time evolution of electronic and vibrational coherences, specifically in conjugated polymers, where strong correlation and population dynamics can affect exciton delocalization \cite{consani2015JChemPhys,banerji2013JMatChemC,yang2005PhysRevB,collini2009Sci,dykstra2009JPhysChemB,nelson2017JPhysChemLett}, vibrational relaxation \cite{heeger2014AdvMat,banerji2011JPhysChemC,yu2012JPhysChemLett,bragg2016JPhysChemLett}, electron transfer \cite{DeSio2016NatComm, DeSio2017PhysChemChemPhys,popp2019JPhysChemLett}, etc. It has been shown that analyzing the 2D spectra as a function of the population time is crucial to estimate and isolate contributions to the homogeneous and inhomogeneous broadening of the spectra due to excited and vibronic states \cite{Scholes2000PhysRevB,Song2014NatComm,song2015JCP}.

Scholes et~al.~\cite{Scholes2000PhysRevB} performed measurements on N,N-bis-dimethylphenyl-2,4,6,8-perylenetetra-carbonyl diamide (PERY) and poly[2-(2'-ethylhexyloxy)-5-methoxy-1,4-phen-ylenevinylene] (MEH-PPV) in toluene as a function of population time. The authors were able to demonstrate the effect on the absorption line-shape due to Coulomb coupling between energetically disordered sub units along the polymer backbone. 

Song et~al.~\cite{Song2014NatComm} reported 2D electronic coherence measurements as a function of population time $\tau_2$ on P3HT:PCBM, a fullerene derivative, nanoparticle blends. The authors found a photoinduced absorption associated to hole formation that appeared at a population time of $\sim 380$\,fs, suggesting that exciton relaxation precedes dissociation in neat P3HT nanoparticles. This is consistent with the delayed photoluminescence findings in neat P3HT reported by us previously~\cite{paquin2011charge,paquin2015multi}. In contrast, with the inclusion of PCBM, hole formation was observed with a time constant of $\sim 24$\,fs, indicating that electron transfer occurs through dissociation of hot excitons. In addition, Song et~al.\ found underlining oscillations with a frequency of $\sim 1506$\,cm$^{-1}$ in the MDCS spectra, corresponding to the vibrational coherence from the P3HT C=C stretching mode. They measured a dephasing time of $\sim 250$\,fs for the vibrational coherence which showed that the electron transfer process to the fullerene domain occurs in a non-equilibrium state (specifically the nuclear configuration and charge distribution). The authors mention the strong amplitude of the vibrational coherence as an indication of similar delocalization of the exciton and hole-polaron given the Huang-Rhys factor would diminish otherwise due to exchange narrowing.

A subsequent study Song et~al.~\cite{song2015JCP} focused on P3HT films instead of nanoparticle blends; the authors reported rephasing and non-rephasing MDCS measurements on %P3HT films 
as a function of the population waiting time $\tau_2$. They developed and proposed a method to identify and separate the signal from vibrational coherences of ground and excited electronic states. Fig.~\ref{fig:song_P3HT} shows the absorptive (real part) of the 2D spectra at a population time of $\tau_2=200$\,fs. The accompanying absorption and PL spectra was used by the authors to identify the features in the 2D map, which are the following: (i) Two diagonal peaks correspond to absorption/emission from electronic-vibrational transitions $S_0^{\nu^0}\rightarrow S_1^{\nu^1}$ at 564\,nm, and $S_0^{\nu^0}\rightarrow S_1^{\nu^0}$ at 614\,nm, where $S_0$ and $S_1$ correspond to the ground and first excited electronic state, respectively, and $\nu^0$ and $\nu^1$ represent the addition of 0 or 1 quantum of a vibrational excitation, respectively. % (as mentioned in section~\ref{x}). 
(ii) Two cross-diagonal peaks are present at the excitation/detection (detection/excitation) wavelengths of 564\,nm/614\,nm (614\,nm/564\,nm) that correspond to stimulated emission. (iii) Two features located at the detection wavelength 674\,nm. The position of these 6 main features are highlighted in Fig.~\ref{fig:song_P3HT} by the crossing of horizontal and vertical dashed lines.

The authors found that all the six features in the 2D spectra have an underlining oscillation with a period $\sim20\,\text{fs}$ ($1519\,\text{cm}^{-1}$). They corroborate this by presenting time traces and their Fourier transforms as a function of the population time (see Fig. \ref{fig:scholes}(b)-(f)). The oscillations correspond to the vibrational frequency of the C=C stretching mode, indicating a coherent interaction with the P3HT electronic states. The Fourier amplitudes of the rephasing and non-rephasing 2D maps (at the frequency $1519\,\text{cm}^{-1}$) were used in conjunction with a four-level model to successfully relate each spectral feature either to ground or excited state vibrational coherences. For example, in the case of features with a excitation wavelength of 614\,nm, the nature of the vibrational coherence changed from excited to ground state depending if the measurement was rephasing or non-rephasing. The analysis resulted in direct evidence of a faster dephasing time for excited state ($\sim 244$\,fs) compared to the ground state ($\sim 297$\,fs) coherence, suggesting that electron transfer in P3HT blends can be mediated via vibrational coupling \cite{Tamjura2011JPhysChemC, Tamura2012JChemPhys, DeSio2016NatComm}.

\section{Frenkel biexcitons}\label{sec:biexcitons}

\begin{figure}
    \centering
    \includegraphics[width=\columnwidth]{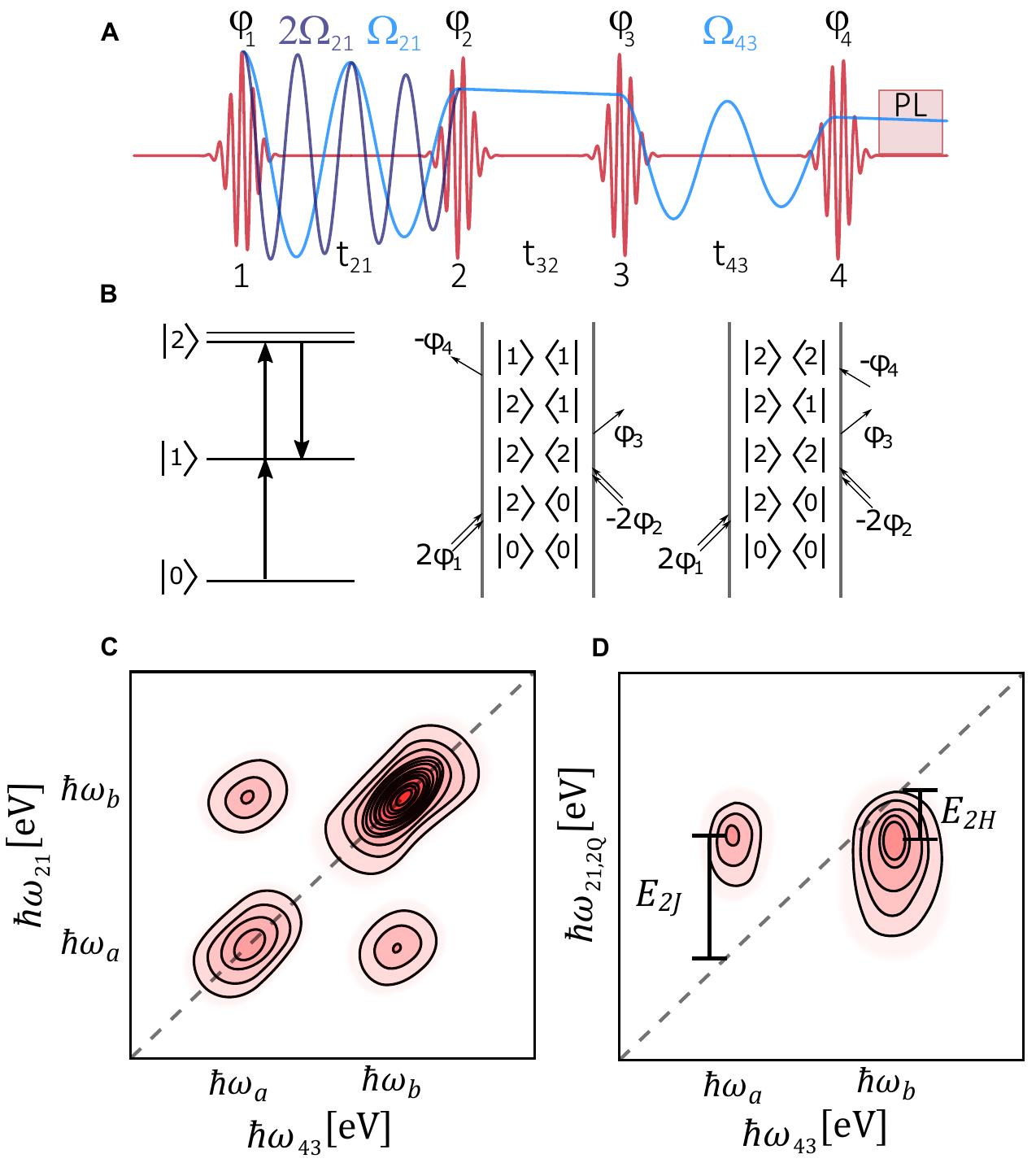}
    \caption{Two-dimensional coherent photoluminescence excitation (2D-PLE) spectroscopy.
        ({\bf \sf A}) Schematic of the experimental pulse sequence. 
    Here $\phi_{i}$ is the phase of pulse $i=1,2,3,4$. The inter-pulse delays are $t_{21}$ (coherence time), $t_{32}$ (population waiting time), and $t_{43}$ (coherence time). The phase-modulation reference waveforms for phase-sensitive detection are generated optically at frequencies $\Omega_{21}$ and $\Omega_{43}$, at which the relative phase $\phi_{21}=\phi_2-\phi_1$ and $\phi_{43}=\phi_4-\phi_3$ oscillate, respectively. In this work, time and spectrally integrated photoluminescence (PL) intensity is demodulated by phase-sensitive detection at the reference frequency $f_{\mathrm{ref}} = \Omega_{43} + \Omega_{21}$ and $ \Omega_{43} + 2\Omega_{21}$ for the one- and two-quantum correlation spectra, respectively, shown in Fig.~\ref{fig:2Q_2D}. We outline in Supplemental Material that the 2D-PLE lineshape can contain contributions from nonlinear incoherent population dynamics over the entire exciton lifetime, and that all spectral lineshapes presented in Fig.~\ref{fig:2Q_2D} are free from this undesired contribution under the excitation conditions of this experiment. 
    ({\bf \sf B}) Double-sided Feynman diagrams of the two most important two-quantum response terms that couple a ground state ($|0 \rangle$), a single exciton state ($|1 \rangle$), and a two-exciton state ($|2 \rangle$). 
    ({\bf \sf C}) Schematic representation of the 2D-PLE expected spectrum for two correlated optical transitions at energies $\hbar \omega_a$ and $\hbar \omega_b$. The spectral axes $\hbar \omega_{21}$ and $\hbar \omega_{43}$ are obtained by Fourier transform of the 2D coherent PL decay function along time variables $t_{21}$ and $t_{43}$ at fixed $t_{32}$. 
    ({\bf \sf D}) Schematic representation of a two-quantum 2D-PLE correlatios spectrum. The two-quantum energy $\hbar \omega_{21,2Q}$ corresponds to the two quantum coherences involving pulses 1 and 2, The diagonal line represents $\hbar \omega_{21,2Q} = 2 \hbar \omega_{43}$. }
    \label{fig:Pulses_Feynman}
\end{figure}
\begin{figure}
    \centering
    \includegraphics[width=0.87\columnwidth]{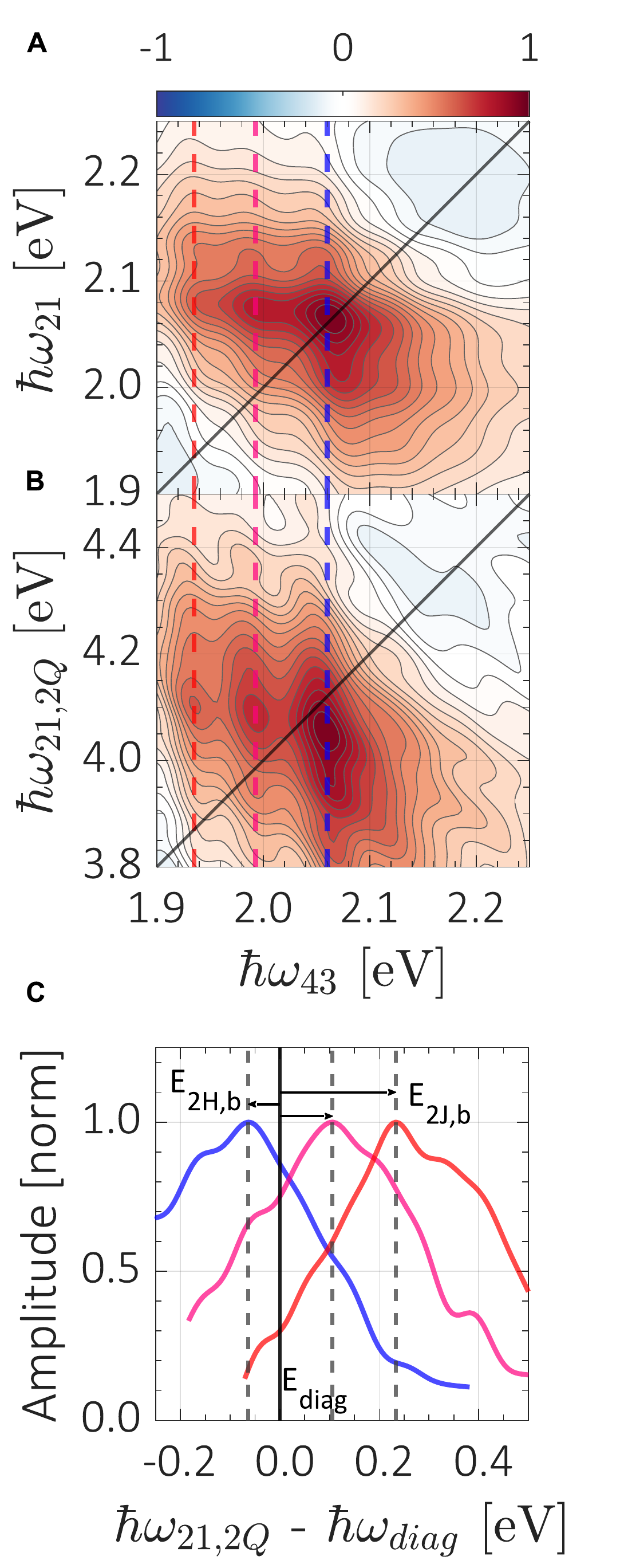}
    \caption{  2D-PLE spectra measured at 5\,K and $t_{32} = 30$\,fs. ({\bf \sf A}) The real part of the one-quantum non-rephasing ($f_{\mathrm{ref}} = \Omega_{43} + \Omega_{21}$) spectrum. ({\bf \sf B}) The real part of the two-quantum ($f_{\mathrm{ref}} = \Omega_{43} + 2\Omega_{21}$)) spectrum.  % ({\bf \sf C}) Diagrammatic representation involving the optical transitions associated with the interpretation of the spectrum in B. 
    ({\bf \sf C}) Spectral cuts along $ \hbar \omega_{21,2Q} - \hbar \omega_{\mathrm{diag}}$ spectral axis at fixed $\hbar \omega_{43} = 2.06$ (blue), $1.99$ (fuscia), and $1.94$\,eV (red) for the spectrum in B. Here $\hbar \omega_{\mathrm{diag}}$ is the $ \hbar \omega_{21,2Q} = 2\hbar \omega_{43}$ two-quantum diagonal energy, corresponding to zero net two-quantum correlation energy (neither binding nor repulsion). We chose to measure 2D coherent spectra with $t_{32}=30$\,fs in order to avoid ambiguous time ordering at $t_{32}=0$. Due to the highly transient nature of the two-quantum coherence signal, it would not be possible to measure with $t_{32}\gg 30$\,fs.}
    \label{fig:2Q_2D}
\end{figure}

Nonlinear coherent spectroscopies can unravel spectral structure beyond excitons as reviewed in Section~\ref{sec:excitons}, but they can also resolve multi-particle binding.
Many-body electronic correlations play a strong role in determining the optical properties of semiconductors~\cite{klingshirn1981optical}. These correlations can give rise to biexcitons, which are bound two-quantum quasiparticles. In organic materials, these two-quantum excitonic states are of fundamental importance, not only because they probe the electronic structure of the material but also because they are expected to be key intermediate states that determine the dynamics of photophysical processes, such as  exciton bimolecular annihilation~\cite{stevens2001exciton}. A direct probe of Frenkel biexcitons in organic systems presents a challenge because of the largely featureless nature and broad inhomogeneous linewidths of Frenkel exciton spectra, in contrast to Wannier-Mott biexcitons in semiconductor quantum wells~\cite{stone2009exciton,turner2010coherent}, which have been well characterized by linear spectroscopy because of well defined spectral structure. An exception to this are Frenkel biexcitons in solid argon~\cite{baba1991formation}. Despite the extensive theoretical work predicting their existence in organic semiconductors~\cite{spano1991biexciton,guo1995stable,gallagher1996theory,agranovich2000kinematic,kun2009effect}, current experimental work in polymeric semiconductors has shown only indirect evidence of two-quantum exciton states by sequential ultrafast excitation~\cite{silva2001efficient,stevens2001exciton,silva2002exciton}. 

%Despite the extensive theoretical work predicting their existence in organic solids \cite{spano1991biexciton,guo1995stable,gallagher1996theory,agranovich2000kinematic,kun2009effect}, hitherto the experimental work in polymeric semiconductors has shown only indirect evidence of two-quantum exciton states by sequential ultrafast excitation  \cite{silva2001efficient,stevens2001exciton,silva2002exciton}. 

\subsection{Two-quantum coherence measurements in PBTTT}
We have recently explored the spectral structure of Frenkel biexcitons in PBTTT, %a conjugated polymer in the polythiophene family, 
with a nonlinear spectral lineshape already discussed in the context of Fig.~\ref{fig:PBTTT_lineshape}, by means of two-quantum coherence measurements~\cite{meza2021molecular}. Such measurements can directly identify biexciton resonances via two-quantum coherences. They can therefore quantify the biexciton binding energies~\cite{stone2009exciton,turner2010coherent}. In that work, we constructed the coherent two-dimensional excitation spectrum via incoherent measurement of the time-integrated photoluminescence (PL) intensity due to a fourth-order excited-state population arising from the interference of wavepackets produced by a sequence of four light-matter interactions~\cite{Tekavec2007JCP}, allowing the measurement of the spectral correlations between resonance involving pulses 1 and 2 and those corresponding to pulses 3 and 4 in Fig.~\ref{fig:Pulses_Feynman}A. This is the same technique as was used for the homogeneous linewidth measurements of P3HT~\cite{Pascal2017PRB}, reported in Fig.~\ref{fig:P3HT_homo}. This technique uses phase sensitive detection to construct the 2D coherent spectrum by modulating the relative phase of each pulse pair in the four-pulse sequence, and demodulating the total photoluminecence signal at a side band of the two phase modulation reference waveforms. However, in Ref.~\citenum{meza2021molecular}, we implemented the demodulation at the second harmonic of the signal relative to the phase modulation function of the first two pulses. The two energy axes are constructed by Fourier transformation of the two-dimensional coherence decay function along time variables $t_{21}$ and $t_{43}$ at a fixed population time $t_{32}$. Accordingly, the spectral correlation along resulting energy axes ($\hbar \omega_{21}$, $\hbar \omega_{43}$) involve single-quantum ($|0\rangle \rightarrow |1\rangle$) and two-quantum  ($|0\rangle \rightrightarrows |2\rangle$) transitions, represented schematically in the left of Fig.~\ref{fig:Pulses_Feynman}B. 
The two principal coherent pathways involving the correlations between one- and two-quantum coherences under this detection scheme are depicted by the double-sided Feynman diagrams in Fig.~\ref{fig:Pulses_Feynman}B. We also display schematic diagrams of the one- and two-quantum 2D-PLE coherent spectra for two correlated transitions in  Figs.~\ref{fig:Pulses_Feynman}C and \ref{fig:Pulses_Feynman}D respectively. These may be correlated, for example, via a common ground state such as H-like and J-like states in a HJ aggregate, each evident in the diagonal of the 2D-PLE spectrum ($\hbar \omega_{21} =  \hbar \omega_{43}$). %In addition, the spectral correlation between these two peaks is manifested by cross peaks between the diagonal features. 
%Similarly, i
In the two-quantum correlation spectrum, a signal along the two-quantum diagonal ($\hbar \omega_{21-2Q} = 2 \hbar \omega_{43}$) signifies two non-interacting excitons, while a signal above or below the diagonal signifies binding interactions with repulsive or attractive character, depending on the signs of the excitonic coupling (H- or J-like coupling)~\cite{spano2014AnnRevPhysChem}, respectively. In %our schematic in 
Fig.~\ref{fig:Pulses_Feynman}D, the lower energy diagonal peak displays 2J biexcitons (the two-quantum energy is higher than twice the one-quantum energy), while the higher-energy resonance displays structure 
corresponding to 2H biexcitons (in which the two-quantum energy is less than twice the one-quantum energy). 

%We first focus on the single-quantum 2D-PLE lineshape in Fig.~\ref{fig:2Q_2D}A. 
The 2D spectrum in Fig.~\ref{fig:2Q_2D}A is dominated by a symmetric diagonal peak centered at $\hbar \omega_{21}=\hbar \omega_{43} \approx 2.06$\,eV, corresponding to the 0--0 peak energy.  
We also observe a weak diagonal signal centered at $\sim 1.99$\,eV, with even weaker structure identified at lower energy. These features display intense cross peaks with the (0--0) resonance. The rich spectral structure displayed at the low-energy tail of the 2D-PLE spectrum %is not evident in the featureless linear absorption spectrum since it is obscured by the inhomogeneous lineshape, and 
demonstrates the existence of distinct states at the low-energy edge of the 0--0 absorption peak. We note that the energy of the weak diagonal feature corresponds to the 0--0 absorbance peak energy found in PBTTT J-aggregates induced when blending this semiconductor with a polar commodity plastic ~\cite{hellmann2013controlling}. 
We thus hypothesize that these are weak signatures of J-aggregate macromolecular conformations bearing effects of interactions with static dipoles.  We present in Ref.~\citenum{meza2021molecular} the linear absorption spectrum of the same batch of PBTTT in a blend with an ionic liquid, processed such that the J-aggregate dominates the lineshape, which supports the assignment in Fig.~\ref{fig:2Q_2D}. The 0--0 absorption peak in that film is at 1.96\,eV, consistent with the spectrum reported in ref.~\citenum{hellmann2013controlling}, which supports the assignment of the weak, low-energy diagonal feature as the origin of the J-aggregate progression. 

%We next examine two-quantum correlations within the spectral structure in Fig.~\ref{fig:2Q_2D}B, looking for exciton-exciton resonances that display repulsive or attractive biexciton binding energies as depicted in Fig.~\ref{fig:Pulses_Feynman}D. 
%Most prominently, w
We observe in Fig.~\ref{fig:2Q_2D}B a broad distribution along the two-quantum energy axis $\hbar \omega_{21,2Q}$ centered at the one-quantum energy $\hbar \omega_{43} = 2.06$\,eV (the 0--0 excitation maximum), with the peak of the distribution below the two-quantum diagonal, \textit{i.e.}, below the energy at which $ \hbar \omega_{21,2Q} = 2\hbar \omega_{43}$. Thereby, the spectrum cuts through the diagonal line, with a tail extending to the high-energy side of the diagonal. We also observe two-quantum peaks for the features at lower energy than the 0--0 peak in Fig.~\ref{fig:2Q_2D}A. These are both centered at higher energy than the diagonal; meaning that the two-quantum correlation is predominantly repulsive for the low-energy features. In Fig.~\ref{fig:2Q_2D}C we display cuts of the two-quantum excitation spectrum at fixed one-quantum energies $\hbar \omega_{43}$. The cuts are  along $\hbar \omega_{21,2Q}$ relative to the two-quantum diagonal energy $\hbar \omega_{\mathrm{diag}}$. This is a good reference because signal on the diagonal corresponds to the energy of two excitons that coexist without interaction. %This then allows us to quantify biexciton binding energies. 
%We discuss the structure in Fig.~\ref{fig:2Q_2D}C in detail in the Supplemental Material. 
Importantly, we notice, on the one hand, that the 0--0 absorption peak, assigned to the vibronic progression of an H aggregate~\cite{spano2014AnnRevPhysChem}, forms biexcitons with 
\textit{attractive} interactions with $E_{\mathrm{2H,b}} = -64 \pm 6$\,meV. %, corresponding to the situation in which $t_{\mathrm{2Q,intra}}<0, U<0$. 
On the other hand, the low-energy resonances, which we hypothesize to correspond to J-aggregate resonances, display predominantly repulsive two-quantum correlations, peaked at $E_{\mathrm{2J,b}} = +106 \pm 6$ and $+233 \pm 6$\,meV, respectively. %, which is produced when $t_{\mathrm{2Q,inter}}>0, U>0$. 
This supports the hypothesis that both attractive and repulsive biexciton binding coexist within the inhomogeneously broadened 0--0 linear absorption peak. 

A key finding of Ref.~\citenum{meza2021molecular} is that that \emph{interchain} H-like excitons are associated with \emph{intrachain} exciton-exciton couplings. In contrast, \emph{intrachain} J-like excitons are paired by \emph{interchain} exciton-exciton couplings. In either case, the biexciton binding energy is related to the exciton-exciton contact interaction and the inter-site hopping (excitonic coupling) energy, which are intrinsic molecular parameters quantified by quantum chemistry and that are expected to be controllable via chemical design and polymer assembly.

\subsection{Biexciton stability and exciton-backbone coupling}
We generally think of bound states as arising from 
and attractive interaction between two particles. 
However, in Ref.~\citenum{meza2021molecular}, as described in the preceeding section, we reported both \emph{attractive} and \emph{repulsive} biexciton binding; we reported further theoretical analysis in Ref.~\citenum{bittner2022concerning}, where we discussed that in J-aggregate systems, 2J-biexcitons can arise from repulsive dipolar interactions with energies $E_{2J}> 2E_J$ while in H-aggregates, 2H-biexciton states $E_{2H} < 2E_H$ corresponding to attractive dipole exciton-exciton interactions. According to our model~\cite{bittner2022concerning}, Frenkel biexcitons mix J-like and H-like character in terms of their collective quantum behavior with the requirement that the ratio of the exciton/exciton interaction and the perpendicular hopping term be $U/t > 0$, with $t = -\hbar^2/2\mu$, $\mu$ the effective mass of the exciton, and $U$ a contact interaction potential, which gives rise to localized biexciton states in the perpendicular direction.  

We an understand this in the context of a textbook-level 
model for a biexciton state interacting with a deformable 
lattice. 
Adopting a relative coordinate between the two excitons, 
we write the biexciton Schr\"odinger equation as 
\begin{align}
    t\psi'' + U\delta(x)\psi = E\psi,
\end{align}
where $U$ is the contact interaction between the two excitons.  
For bound states, $\psi(x)$ must vanish as $x\to\pm\infty$, giving that
\begin{align}
    \psi(x) = \left\{
    \begin{array}{cc}
      \sqrt{\kappa}e^{-\kappa x}   &  x > 0\\
       \sqrt{\kappa}e^{+\kappa x}  & x < 0, 
    \end{array}
    \right.
    \label{eq:20}
\end{align}
where $\kappa = U/2t$ is a positive constant and $E = t \kappa^2$. In general, we take 
$t = -\hbar^2/2\mu_{eff}$ and $U<0$ for an attractive potential giving rise to a bound state
energetically {\em below} the continuum for the scattering states.  

Generally, interactions with the lattice phonons 
produce an additional stability to to the reorganization
of the system about the bound state.  However, in the 
case of the 2J bi-exciton, reorganization of the 
lattice can actually {\em destabilize} the state by 
pulling it back into the continuum of unbound states. 
To see this, we append to the 1D impurity model a term coupling the biexciton to the lattice as per the Davydov model, \cite{SCOTT19921,Edler2004DirectOO,PhysRevB.82.014305,doi:10.1063/1.3592155,Goj:10,GEORGIEV2019257,GEORGIEV2019275,PhysRevB.35.3629,DAVYDOV1977379,DAVYDOV1973559}
such that resulting equations of motion read
\begin{align}
i\hbar \dot \psi(x) &=  (t \nabla^2  + U\delta(x) + (E_o + 2 \chi \nabla u(x))) \psi(x) \nonumber \\ 
\ddot u - \frac{k}{m}\nabla^2 u &= 2\frac{\chi}{m}\nabla|\psi|^2,
\end{align}
where $u(x)$ is the lattice deformation, $\chi$ is the linear coupling between the biexciton and the lattice, 
$m$ is the 
mass of the lattice ``atoms'' 
and $k$ is the elastic modulus. 
The bound-states are invariant under 
Galilean transformation and one can 
find a closure relation
\begin{align}
\nabla u = -\frac{2\chi}{k} | \psi|^2
\end{align}
that gives us a non-linear Schr\"odinger equation
\begin{align}
i\hbar \dot \psi = (t\nabla^2 + g|\psi|^2  +U\delta(x))\psi,
\end{align}
where $g = - 4\chi^2/k$.
Note that $E_0$ is a constant given by 
\begin{align}
    E_0 = E - 2t +\frac{1}{2}
    \int_{-\infty}^{\infty}
    \left( m \ddot u^2 + k u'' \right)dx
\end{align}
that we can ignore for purposes of this analysis.
The $\delta$-function potential implies the wave function 
should have the form in Eq.~\ref{eq:20}. 
Taking $\kappa$ as a variational 
parameter and minimizing the 
total energy, one obtains
\begin{align}
\kappa = \frac{U}{2t} + \frac{g}{8t}.
\end{align}
 $\kappa > 0$ is necessary to produce a localized state
 and from above  $U/t > 0$  and $g<0$ from its definition above, 
we have a stability requirement that if $U>0$ and $t>0$, then $-g< 4U$. 
Solving for the binding energy, 
\begin{align}
E_B = \frac{(4U + g)^2}{64 t}.
\end{align}
We obtain a straight-forward estimate of the contribution of both the 
lattice and the exciton/exciton coupling to the biexciton binding. 

\begin{figure}
\includegraphics[width=\columnwidth]{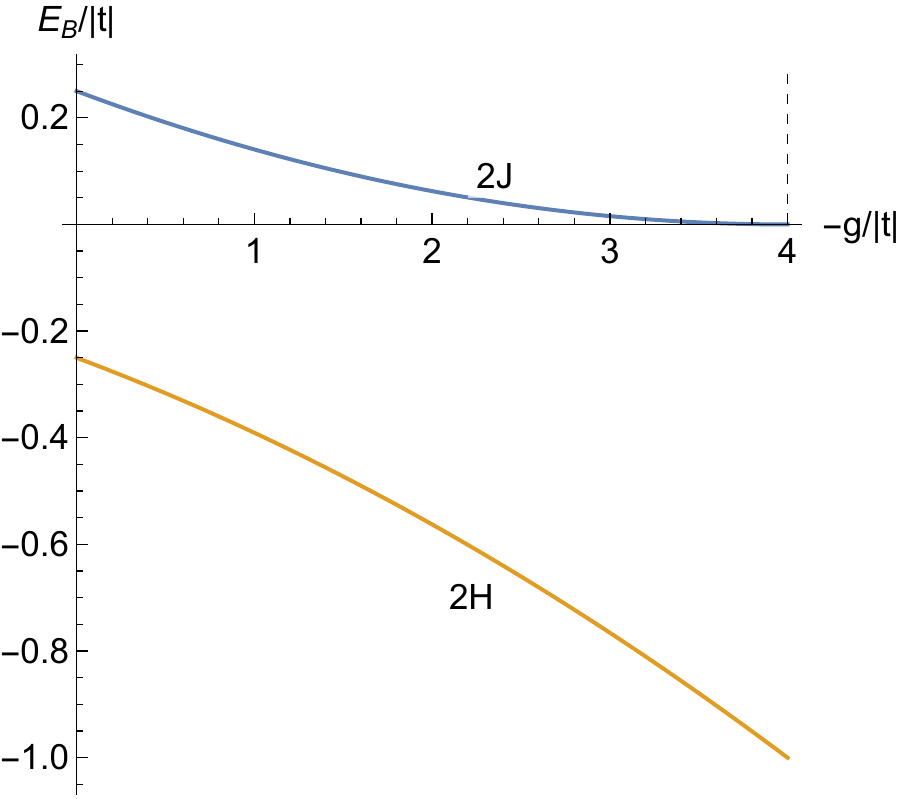}
\caption{Biexciton binding energy (in reduced units) for the 2J and 2H cases versus 
increasing exciton/lattice coupling.   For the attractively bound 2H, lattice reorganization
is expected to stabilize the biexciton state while destabilizing the 2J.
The dashed line indicates the limit of stability for the 2J case.}
\label{fig:stab}
\end{figure}
\begin{figure*}
\centering
\includegraphics[width=0.70\textwidth]{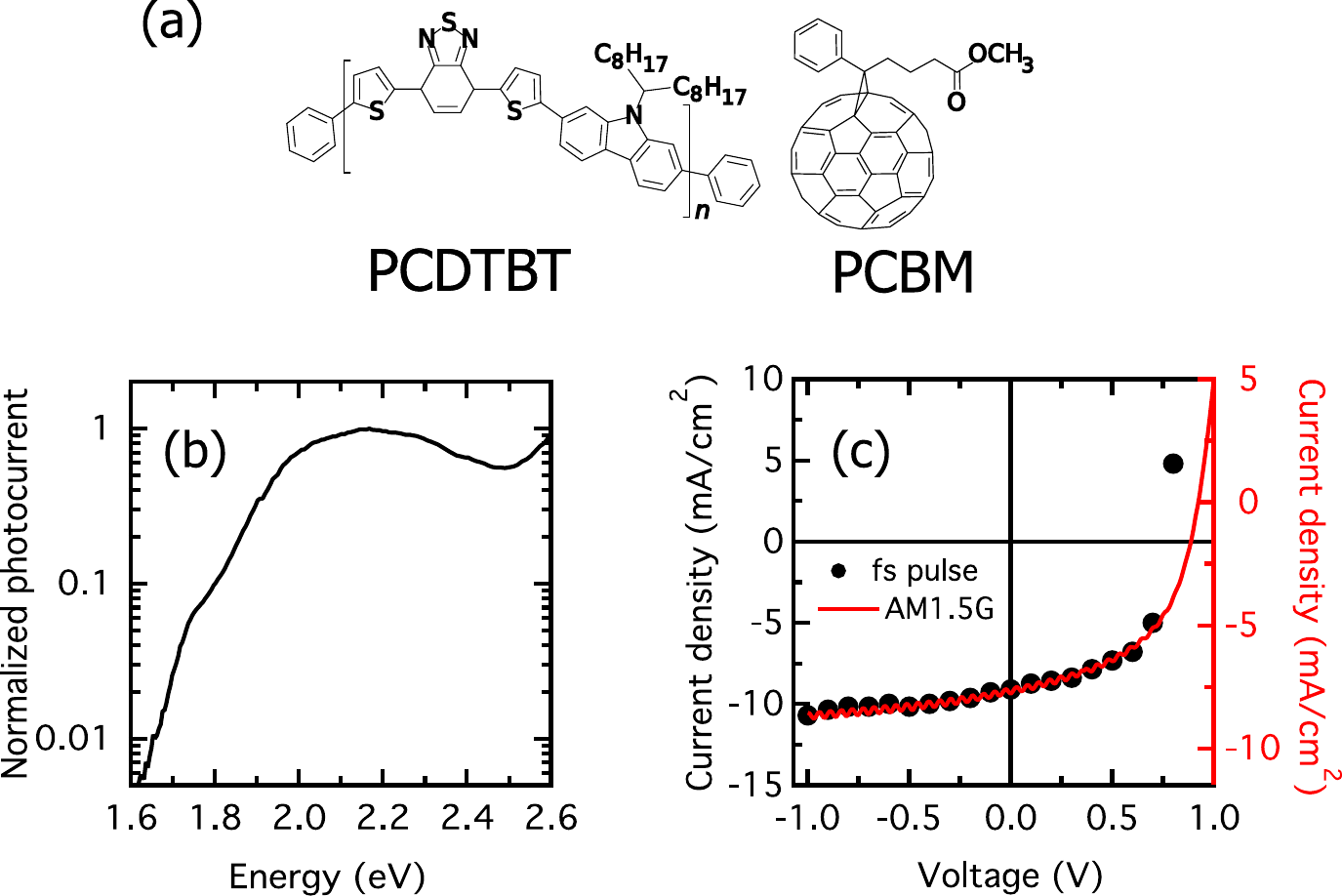}
\caption{(a) Structure of PCDTBT (left) and PCBM (right). (b) Photocurrent excitation spectrum of the PCDTBT:PCBM solar cell under investigation.  (c) Current-voltage response of same cell under AM 1.5\,G solar illumination (right-axis) and under laser illumination (left-axis). Figure extracted from Ref.~\citenum{vella2016ASciRep}.}
\label{fig:JV_PLE}
\end{figure*}

\begin{figure}
\centering
\includegraphics[width=0.7\columnwidth]{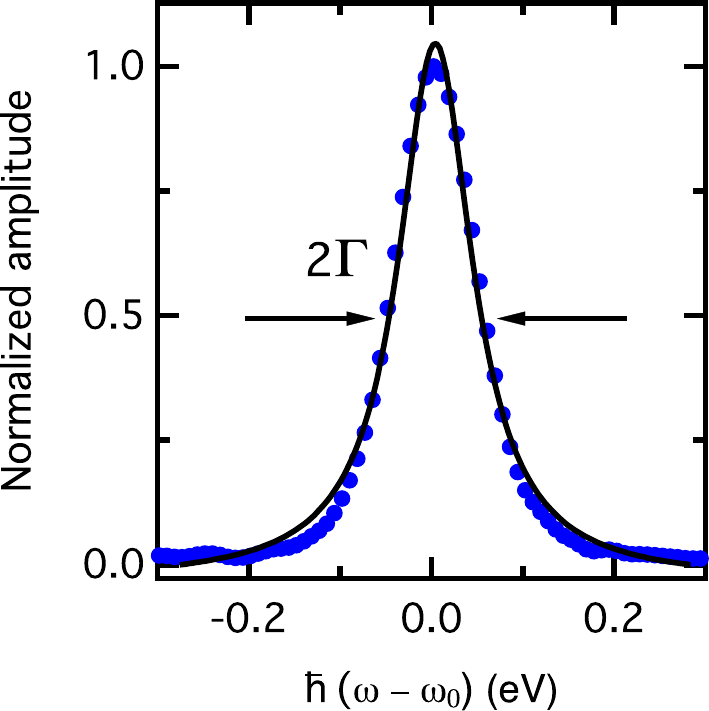}
\caption{Anti-diagonal slice of the 2D-PCE rephasing signal. The solid line is a fit to a Lorentzian function. % in which the optical dephasing rate $\Gamma$ characterizes the homogeneous linewidth. 
Figure extracted from Ref.~\citenum{vella2016ASciRep}. }
\label{Lorentzian_profile}
\end{figure}

In Fig.~\ref{fig:stab} we plot the biexciton binding energy versus the non-linearity parameter, $g$.
For the attractively bound 2H, lattice reorganization
is expected to stabilize the biexciton state by further localizing the 
state ($\kappa$ increases as $g$ increases in magnitude).  On the other hand, 
for the 2J state, increasing the magnitude  of $g$ decreases $\kappa$ and 
\textit{destabilizes}
the otherwise bound 2J state by effectively pulling it back
into the continuum of unbound (scattering) biexciton 
states.
When $-g = 2U$ the state is fully delocalized and
further increases in the lattice coupling lead to unbound solutions.  We rationalize the biexciton spectral structure observed in Fig.~\ref{fig:2Q_2D} by the dependence of the biexciton binding energy on $g$ as reported in Fig.~\ref{fig:stab}.

\section{Photocurrent-detected spectroscopy in operating diodes}

Two-dimensional photocurrent excitation spectroscopy (2D-PCE) is an MDCS experiment, where the detection mechanism probes the photocurrent generated by a sequence of four ultrashort laser pulses, akin to the photoluminescence-detected variant of the experiment in Fig.~\ref{fig:2Q_2D}. 2D-PCE has been a key approach in the research of photocharge generation dynamics, not only because of the unique detection sensitivity but also due to the ability to probe the correlation between optical excitations and photocarriers in polymer semiconductors~\cite{vella2016ASciRep}. These correlations are identifiable through cross-peaks in 2D spectral maps, and are of the utmost importance to study materials in which photocarrier precursors, such as excitons in molecular semiconductors, are the primary photoexcitations. A special case are materials for photovoltaic applications, such as solar cells based on PCDTBT:PCBM blends (Fig.~\ref{fig:JV_PLE}(a)). This is a polymer:fullerence benchmark system that has been shown to yield solar-power conversion efficiencies as high as 7\%~\cite{park2009bulk,sun2011efficient,vella2016ASciRep}. The photocurrent excitation spectrum of this system is diaplayed in Fig.~\ref{fig:JV_PLE}(b). 

%is a direct probe of the couplings between different excited states through the presence of crosspeaks (off-diagonal spectral features) which is of the utmost importance to study materials in which photocarrier precursors such as excitons in molecular semiconductors, are the primary photoexcitations.

%Correspondingly, Li et. al. \cite{li2016probing} reported 2DPCE measurements on a solar cell based on PCDTBT:PCBM blend, which is a polymer:fullerence benchmark system that has been shown to yield solar-power conversion efficiencies as high as 7\% \cite{park2009bulk,sun2011efficient}.
%(using JV curves)
%In a previous work regardig PCDTBT-PCBM blends, this polymer system, 
Importantly, Vella et~al.~\cite{vella2016ASciRep} demonstrated that %an open-circuit voltage is similar whether you illuminate 
illumination using an ultrafast pulse sequence such as that depicted in Fig.~\ref{fig:2Q_2D}, at sufficiently low fluence, produces J-V curves that match those under standard solar (AM 1.5G) illumination, %. This is 
shown in Fig.~\ref{fig:JV_PLE}(c), where the current density (as a function is applied voltage) is compared between both types of illumination. This result suggests that the current densities probed by 2D-PCE experiments are comparable to those of a functioning solar cell, %(see Fig. \ref{fig:JV_PLE} b).
which is critical in establishing that these advanced optical probes probe dynamics in regimes that are relevant in the optoelectronic device. 

Additionally, as a result of 2D-PCE measurements, Vella et~al.\ suggested %that photocurrent detection was evidence of 
a tunneling process from an exciton state to a current producing charge-separated state. Consequently, the authors stated that the timescale of this process would be dominated by fluctuations due to interaction with the bath ensemble, predicting dephasing times $\leq 20$\,fs. In order to test their prediction, the authors extracted the optical dephasing time by taking the antidiagonal cut of the 2D spectra, as shown in  Fig.~\ref{Lorentzian_profile}. The measured peak was fit to a Lorentzian function with a full-width-at-half-maximum (FWHM) of 100$\pm 1$\,meV. This represents the homogeneous linewidth of the main exciton transition from the donor polymer, and corresponds to a dephasing time of 13\,fs~\cite{vella2016ASciRep}. 

\begin{figure}
   \centering
    \includegraphics[width=0.9\columnwidth]{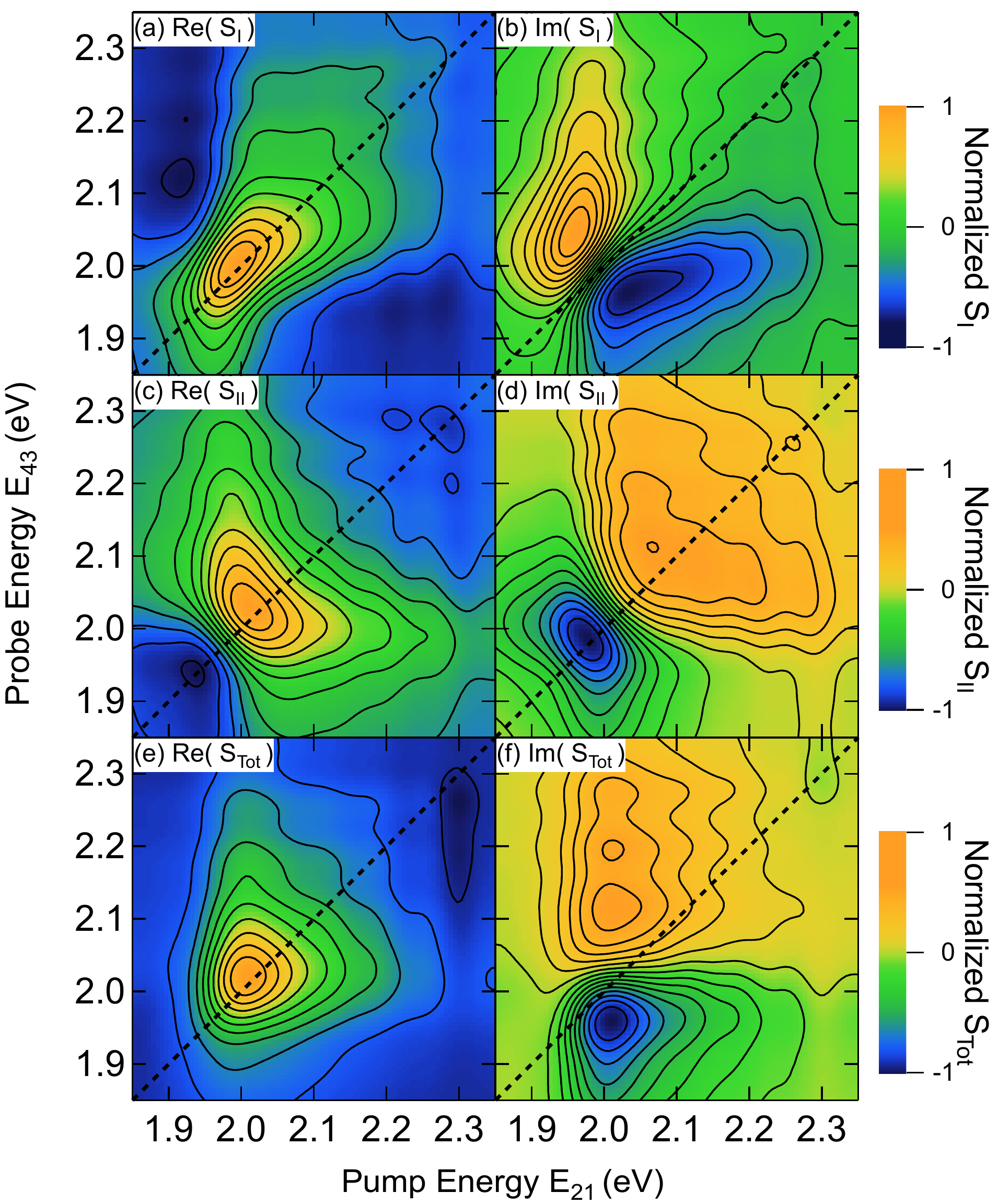}
    \caption{2D PC measurements on a PCDTBT:PCBM photovoltaic diode (see Fig.~\ref{fig:JV_PLE}) at ambient temperature. Shown are the real (left column) and imaginary (right column) parts of the rephasing (a, b), non-rephasing (c,d), and total correlation function (e,f). The population time is 50\,fs. Figure extracted from Ref.~\citenum{li2016probing}.}
    \label{fig:PCDTBT_PC}
\end{figure}

A concurrent 2D-PCE study by Li et~al.~\cite{li2016probing} %performed  measurements on 
on PCDTBT:PCBM, similar to that of Vella et~al.~\cite{vella2016ASciRep}, but with larger bandwidth of the ultrafast pulse,  %; the experiments 
revealed an inhomogeneously broadened spectrum, %along the diagonal, indicating 
and suggested structure of charge-transfer excitons and %exciton produced 
photocarriers. Fig.~\ref{fig:PCDTBT_PC} was extracted from this study, and it shows the 2D maps corresponding to the rephasing (a,b), non-rephasing (c,d) and total spectrum (e,f). In addition, the authors demonstrated that the 2D spectra from photocurrent excitation (2D-PCE), photoluminescence excitation (2D-PLE), and photoinduced absorption (2D-PIA) all yield similar features, indicating that charge-transfer excitons, probed via photoluminescence, photocarriers via photocurrent, and polarons generally via quasi-steady-state photoinduced absorption, can all be probed using the same excitation pulsetrain sequence. This work further confirmed the polaronic nature of the photocarriers and charge-transfer excitons in PCDTBT:PCBM.

\section{Perspective}

Given the work that we have highlighted in this review, in this section we provide out vision of where MDCS can make further impact in the understanding of excitons in $\pi$-conjugated polymers. 

\subsection{Coherent optical lineshapes of photophysical aggregates}\label{sec:persp_lineshape}
   
Besides disentangling the homogeneous and inhomogeneous contributions to the total spectral width, and thereby quantifying the optical dephasing processes in materials, MDCS can reveal a very well defined spectral structure as shown in  Fig.~\ref{fig:PBTTT_lineshape}, which displays the absolute, real, and imaginary spectra of the rephasing signal of a PBTTT film at 5\,K and at a population waiting time $\tau_2 = 0$\,fs. If we observe carefully Fig.~\ref{fig:PBTTT_lineshape}(b) and Fig.~\ref{fig:PBTTT_lineshape}(c), %which show the real and imaginary components acquired at zero population waiting time, 
%the diagonal 0--0 and 0--1 features at energies $\sim 2.06$ and $\sim 2.21$\,eV, evolve with different absorptive and dispersive character. This behavior is expected given the intrinsic photophysical character difference between 0--0 and 0--1. The 0--0 transition is sensitive to the exciton coherence\cite{spano2009determining} while 0--1 is mainly incoherent.
The real part of the 0--0 peak at $\sim 2.06$\,eV displays absorptive lineshape in the real component, but dispersive character in the imaginary component; a diagonal peak in the real part and a derivative feature about the diagonal in the imaginary part. This is what is expected for non-interacting chromophores. However, the 0--1 diagonal feature at $\sim 2.21$\,eV displays inverted behaviour: the real part displays apparently dispersive lineshape while the imaginary part displays apparently dispersive lineshape. This behaviour evolves over the first few hundred femtoseconds (data not shown here) such that this character inverts between 0--0 and 0--1 diagonal peaks: 0--0 in the real part evolves from absorptive to dispersive character in the real spectrum and vice-versa in the imaginary one. This behaviour encodes the quantum dynamics of the photophysical aggregates. It is well understood now that the 0--0 feature encodes the many-body behaviour of the aggregate~\cite{Spano2005JChemPhys}. It will be highly insightful to describe this type of MDCS lineshape evolution by simulation of the quantum dynamics that include all of the aggregate many-body details, lattice reorganization, and energetic disorder. Our groups are currently focused in this work. 

It was also highlighted in Section~\ref{sec:PBTTT_lineshape} that the absolute spectrum in Fig.~\ref{fig:PBTTT_lineshape}(a) display highly symmetric diagonal 0--0 and 0--1 diagonal peaks. This reflects a comparable values of the homogeneous and inhomogeneous linewidths, which we interpret as indicative of dynamic disorder governing the disordered landscape in PBTTT. On the other, P3HT is clearly in an inhomogeneously broadened limit, as shown in Fig.~\ref{fig:P3HT_homo}. The linear absorption spectrum of both materials, however, displays similar total width, and it would not have been possible to differentiate between the two materials purely from linear spectroscopy. This presents an opportunity to investigate more technologically relevant families of conjugated polymers that have varying degree of backbone rididity, and that span from flexible polymers such as P3HT to ``hairy-rod'' materials~\cite{balzer2022structure}, for example.  

Another opportunity for MDCS is to address lineshapes in push-pull materials~\cite{chang2021hj,zhong2020unusual,balooch2020vibronic}. There are already reports of these techniques applied to this class of materials~\cite{Song2019Multispectral,Song2021Mechanistic}. MDCS approaches are well suited to address the structure of inter-chromophore coupling effects due to charge-transfer interactions, and we consider that there is still much to understand in structure-property relations in this realm. 

\subsection{Two-quantum spectral structure of Frenkel biexcitons }

The rich spectral structure revealed in Fig.~\ref{fig:2Q_2D} is intriguing in that it reveals the role of both attractive and repulsive binding in conjugated polymer hybrid HJ aggregates. We rationalized this observation by considering the geometry of two-quantum correlations along and between chains, but there is tremendous scope to understand this further with microscopic detail, and we consider that there is substantial scope for the electronic structure community to push the development of first-principle methods that account for many-body correlations to explore our reports of biexciton binding in PBTTT~\cite{meza2021molecular}. On the same token, stochastic effects that drive many-body quantum dynamics~\cite{li2022optical} need to be incorporated in the theoretical treatment of Frenkel biexcitons.  Finally, we consider that the series of materials families that will be relevant for 2D coherent lineshape analysis in the context of the discussion in Section~\ref{sec:persp_lineshape} should be studied with the double-quantum coherence techniques developed in Ref.~\citenum{meza2021molecular}, and the measurement should be executed beyond two-quantum correlations --- how many Frenkel excitons correlate in the series of conjugated polymers of technological interest, and what does this imply for their potential in classical and quantum photonics?

\subsection{MDCS in operating optoelectronic devices}
 
Fig.~\ref{fig:JV_PLE} addresses a general criticism of ultrafast spectroscopy, which affirms that the excitation conditions and resulting excitation densities are beyond the regimes that are relevant in the operation of polymer-based optoelectronic devices. With photocurrent detection and using the phase-sensitive detection techniques developed originally by Tekavec et~al.~\cite{Tekavec2007JCP}, it is possible to measure nonlinear responses of the polymer semiconductor in a diode with current-voltage characteristics that are in the regime of solar power conversion~\cite{vella2016ASciRep}. We consider that there is ample scope to expand the studies to address contemporary high-efficiency systems involving non-fullerene acceptor systems, in which the spectral structure of interchromophore couplings will play a critical role in the understanding of device operation. One unexplored direction is in electrochemical cells. We have been able to carry out incoherent ultrafast experiments in conjugated-polymer-based electrochemical cells~\cite{bargigia2021charge}, and there are clear opportunities to extend such studies with the photocurrent/photovoltage spectrsocopies discussed in this review. Nevertheless, the role of unexpected incoherent nonlinear population dynamics may be an issue in organic devices~\cite{Gregoire2017,Bargigia_GaAs}, and methods to decouple this background signal from the nonlinear coherent response will be required. However, this also presents an opportunity to develop methods to measure concurrently nonlinear population dynamics between multiple excitations, and the nonlinear coherent response of the material.

%PBTTT is an interesting material owing its thermotropic phase behavior which give rise to a very dynamic disordered energy landscape.   

%   \begin{itemize}
 %   \item 2Q -- Frenkel biexcitons?
  %  \item lineshape analysis for many-body signatures
%\end{itemize}    
    
   % \item dynamic versus static disorder
      %The absorption and photoluminescence spectra of PBTTT, is consistent with the hybryd HJ-aggregate model (include ref). From the linear spectrum shown in Fig.~\ref{fig:PBTTT_Absrephzerotime} we observe a dominant H-like character.
    %Real and imaginary components  of the rephasing signal taken at different waiting population times, reveal important information on the possible aggregation effects in the exciton dynamics. 
    %figuras 11 y 12

%%%END OF MAIN TEXT%%%


\balance


\section*{Author contributions}
EGM and AVF gathered most of the material described in this review article, and led the early drafts of the manuscript. CSA and ERB led the intellectual development of the article. All authors participated in the redaction of the manuscript. 

\section*{Conflicts of interest}
There are no conflicts of interest to declare by any of the authors. 

%%%ACKNOWLEDGEMENTS%%%
\section*{Acknowledgements}

The work at Georgia Tech was funded by the National Science Foundation (DMR-1904293). CS acknowledges support from the School of Chemistry and Biochemistry and the College of Science at Georgia Tech. The work at the University of Houston was funded in part by the National Science Foundation (CHE-2102506, DMR-1903785) and the Robert A. Welch Foundation (E-1337).

%%%REFERENCES%%%
%\bibliography{rsc} %You need to replace "rsc" on this line with the name of your .bib file
\bibliographystyle{rsc} %the RSC's .bst file
\providecommand*{\mcitethebibliography}{\thebibliography}
\csname @ifundefined\endcsname{endmcitethebibliography}
{\let\endmcitethebibliography\endthebibliography}{}

\end{document}